%% file: main.tex
\documentclass[fleqn,10pt]{wlscirep}
\usepackage[utf8]{inputenc}
\usepackage[T1]{fontenc}

\usepackage{comment}
\usepackage{mathtools}

\usepackage{stackengine}
\usepackage{multirow}
\usepackage{textcomp}
\usepackage[draft]{todonotes}
\usepackage{makecell}
\usepackage{subcaption}

\usepackage{graphicx}

\usepackage{xspace}
\input{min_macros}

\input{rs_macros_general}

\newcommand{\JSChange}[1]{{\color{black} {#1}}}

\newcommand{\xsk}{\ensuremath{\underline{\mathbf{x}^k}}\xspace}
\newcommand{\ask}{\ensuremath{\underline{\mathbf{a}^k}}\xspace}

\usepackage[normalem]{ulem}
\usepackage{braket}
\def\rot{\rotatebox}
\def\imp{\rightarrow}

\title{Effective Prime Factorization via Quantum Annealing
by Modular Locally-structured Embedding}

\author[1]{Jingwen Ding}
\author[1]{Giuseppe Spallitta}
\author[1]{Roberto Sebastiani}
\affil[1]{University of Trento, Trento (Italy)}

\affil[1]{\{jingwen.ding, giuseppe.spallitta, roberto.sebastiani\}@unitn.it}

\begin{abstract}
  This paper investigates novel techniques to solve prime
  factorization by quantum annealing (QA). 
  Our contribution is twofold.

  First,  we present a novel and very compact modular {\em encoding} of a binary multiplier circuit into
  the Pegasus architecture of current D-Wave QA devices.
  The key contribution is a compact encoding of a controlled full-adder
into an 8-qubit module in the Pegasus topology, which we synthesized offline by means of
Optimization Modulo Theories.
This allows us to encode up to a 21×12-bit multiplier (and a 22×8-bit one)
into the Pegasus 5760-qubit topology of current annealers. 
To the best of our knowledge, these are the largest
  factorization problems ever encoded into a quantum annealer.

    Second, we have
  investigated the problem of actually {\em solving} encoded PF problems
  by running an extensive experimental evaluation on
  a D-Wave Advantage 4.1 quantum annealer.
   In order to help
the annealer in reaching the global minimum, in the
experiments we introduced different approaches to initialize the
multiplier qubits 
and adopted several 
performance enhancement techniques.
Overall, exploiting all the encoding and solving techniques described
in this paper, 
{$8,219,999=32,749 \times 251$}
was the highest prime product we were
 able to factorize within the limits of our QPU resources.
 To the best of our knowledge, this is the largest number which
 was ever factorized by means of a quantum annealer, {and, more generally, by a quantum device}.

  \ignore{  
  This paper investigates novel techniques to solve prime
  factorization by quantum annealing. First, we focus on encoding
  multipliers on currently available quantum annealers architecture,
  by proposing a modular approach that minimizes the amount of qubits
  necessary for each controlled full adder (CFA). In particular, we
  propose several techniques, such as alternating CFAs, qubit sharing,
  and virtual qubit chaining, that allow different full adders to
  partially overlap, using the same qubits on different CFAs with
  different roles.  By combining all these techniques we encoded a
  $21\times12$ bits multiplier on a Dwave Pegasus 4.1 system, the
  biggest multiplier ever encoded on a Pegasus architecture. 
  
  Second, taking into account current hardware limitations (faulty
  qubits/coupling and limited QPU time) we propose several approaches
  to solve gradually bigger prime factorization problems. These
  techniques include more efficient initialization of some qubits'
  value, thermal relaxation during forward annealing, and the
  integration of reverse annealing to improve sub-optimal
  samples. Considering all these solving techniques we solved prime
  factorization problems up to $15\times8$ bits numbers including
  8273179, the biggest number to be factorized using quantum annealers
  to our knowledge.
  }
  \ignore{
For benchmarking quantum annealing on prime factorizations,
we propose a modular encoding approach from multiplier
based on the locally structured embedding.
This approach can generate a penalty function,
which corresponds to the underlying Ising model
the D-Wave quantum system finally evolves to:
with the minimal gap locally maximized 
through satisfiability/optimization modular theory solvers
and unchanged when the multiplier size increases; 
the number of required qubits and the length of chains 
scales as $\mathcal{O}(mn)$ and $\mathcal{O}(max(m, n))$ respectively wrt a $m\text{bit}\!\times\!n\text{bit}$ multiplier.
It is expected more efficient and scalable than the global-embedding based approaches.
To adapt to the D-Wave Advantage system, the qubit sharing trick is proposed
to further improve the efficiency and scalability of the encoding approach.
It manifests itself in saving qubits for OMT encoding as ancillary variables
and in replacing chains if they are missing in the system.
As a result, it allows us to encode up to a $21\text{bit}\!\times\!12\text{bit}$ multiplier 
so that a faulty-free annealer can be fed an prime factorization
problem up to $8,587,833,345=2,097,151\!\times\!4,095$.
In practice, this maximal embeddable size is reduced to $17\text{bit}\!\times\!8\text{bit}$
due to faulty qubits and couplings 
and the maximal solvable size is further reduced to $14\text{bit} \times 8\text{bit}$
due to the confined annealing time.
Assisted by thermal relaxation and quantum local search,
D-Wave Advantage 4.1 successfully factored some integers 
up to $14\text{bit} \times 8\text{bit}$:
$1, 480, 867 = 8, 273 \times 179$ is an instance of this size factored with non-groundstates,
$378, 227 = 2, 113 \times 179$ is an instance of $12\text{bit} \times 8\text{bit}$
factored with exact groundstates.
}

\end{abstract}
\begin{document}

\flushbottom
\maketitle
%
%
\thispagestyle{empty}

\input{Intro}

\input{foundations}

\input{encoding}

\input{updated_results}

\input{solving}

\input{IRV_improvement}

\section*{Conclusions and Future Work}

In this paper we have proposed a novel approach to prime factorization by quantum
annealing. Our contribution is twofold.

First, we have presented a novel modular {\em encoding} of a binary multiplier circuit into
the Pegasus architecture of the most recent D-Wave QA devices. The key to
success was a compact encoding of a controlled full-adder sub-circuit
into an 8-qubit module in the Pegasus topology, which we synthesized offline by means of
Optimization Modulo Theories. 
This allows us to encode up to a 21×12-bit multiplier (resp. a
22×8-bit one) into a
5760-qubit Advantage 4.1 annealer.
To the best of our knowledge, these are the largest
factorization problems ever encoded into a quantum annealer.
Also,  due to the modularity of the encoding, this number will scale up automatically with the growth of the
  qubit number in future chips. Thus, we believe that this encoding can be
  used as a baseline for many future research for prime factorization
  via QA.

  Second, we have
  investigated the problem of actually {\em solving} encoded PF problems
  by running an extensive experimental evaluation on
  a D-Wave Advantage 4.1 quantum annealer.
Despite the presence of faulty qubits and couplings and within the limited amount of QPU
time we had access to, 
by  exploiting all the encoding and solving techniques we introduced
and described
in this paper, 
{$8,219,999=32,749 \times 251$}
was the highest prime product we were
able to factorize.
 To the best of our knowledge, this is the largest number which
 was  ever factorized by means of a quantum annealer, and
   more generally by a quantum device, without adopting hybrid
 quantum-classical techniques.
 We are confident that 
even better results  can be obtained with a less-faulty annealer and a larger availability of QPU time.

There is still much room for further developments.
First, efficient encodings for alternative multiplier schemata could be developed.\cite{mengoni2020breaking}
Second, other solving strategies within the annealing process could be conceived
and empirically investigated. 
Moreover, D-Wave recently announced the upcoming generation of quantum
processors built on top of a new topology, Zephyr, that provides more
connections and cliques among different sets of qubits.
Once we have access to a large-enough Zephyr processor,
we plan to
test out encoding algorithms to get better penalty functions for the
CFAs and reach global minima more easily during the solving phase.

\newpage

\section*{Data availability}
 Data about the experimental section, and in particular
the code to replicate the solving experiments is publicly accessible here: \url{https://gitlab.com/jingwen.ding/multiplier-encoder/}.
\bibliography{references}

\newpage

\end{document}

%% file: min_macros.tex
\newcommand{\offset}{\ensuremath{{\sf o}}}
\renewcommand{\offset}{\ensuremath{\theta_{0}}}

\newcommand{\xs}{\ensuremath{\underline{\mathbf{x}}}\xspace}

\newcommand{\zs}{\ensuremath{\underline{\mathbf{z}}}\xspace}

\newcommand{\as}{\ensuremath{\underline{\mathbf{a}}}\xspace}

\newcommand{\ts}{}
\renewcommand{\ts}{\ensuremath{\underline{\boldsymbol{\theta}}}\xspace}
\newcommand{\Fx}{\ensuremath{F(\xs)}\xspace}

\newcommand{\Px}{\ensuremath{P_F(\xs|\ts)}\xspace}
\newcommand{\Pxprime}{\ensuremath{P'_F(\xs|\ts)}\xspace}
\newcommand{\Pxa}{\ensuremath{P_F(\xs,\as|\ts)}\xspace}
\newcommand{\Pxak}{\ensuremath{P_{F_k}(\xs^k,\as^k|\ts^k)}\xspace}
\newcommand{\Pxai}{\ensuremath{P_{F_i}(\xs^i,\as^i|\ts^i)}\xspace}

\newcommand{\Pxaj}{\ensuremath{P_{F^{(j)}}(\xs,\as^{(j)}|\ts^{(j)})}\xspace}
\renewcommand{\Pxaj}{\ensuremath{P_{F_j}(\xs^j,\as^j|\ts^j)}\xspace}

\newcommand{\ignoreinshort}[1]{}

%% file: rs_macros_general.tex
\input{rgb}

\newcommand{\TODO}[1]{{}}

\newcommand{\ignore}[1]{}

\newcommand{\RSTODO}[1]{{\bf \textcolor{darkgreen}{{\fbox{RS TODO:} #1}}}}
\renewcommand{\RSTODO}[1]{}

\newenvironment{rschange}{\color{blue}}{\normalcolor}

 \newcommand{\ignoreinlong}[1]{{#1}}
 \renewcommand{\ignoreinshort}[1]{\textcolor{blue}{#1}}
 \renewcommand{\ignoreinlong}[1]{}

\def\makenewenumerate#1#2{%
\newcounter{cnt#1}
\newenvironment{#1}%
{\begin{list}{\makebox[0pt][r]{#2}}%
{\setlength{\itemsep}{0pt}%
 \setlength{\parsep}{.2em}%
 \setlength{\leftmargin}{1.5em}%
 \setlength{\labelwidth}{.4em}%
 \usecounter{cnt#1}}}
{\end{list}}}

\makenewenumerate{myenumerate}{\arabic{cntmyenumerate}.}

\makenewenumerate{renumerate}{\rm(\roman{cntrenumerate})}
\makenewenumerate{renumerateprime}{\rm(\roman{cntrenumerateprime}$'$)}
\makenewenumerate{renumeratesecond}{\rm(\roman{cntrenumeratesecond}$''$)}

\makenewenumerate{aenumerate}{({\it\alph{cntaenumerate}})}
\makenewenumerate{aenumerateprime}{({\it\alph{cntaenumerateprime}$'$})}
\makenewenumerate{aenumeratesecond}{({\it\alph{cntaenumeratesecond}$''$})}

\def\newplaintheorem#1#2{%
\newtheorem{#1plain}{#2}
\newenvironment{#1}{\begin{#1plain}\rm }{\end{#1plain}}}

\newplaintheorem{notation}{Notation}
\newplaintheorem{cexample}{Counter-Example}

\newcommand{\tuple}[1]{\ensuremath{\langle{#1}\rangle}\xspace}
\newcommand{\set}[1]{\ensuremath{\{{#1}\}}\xspace}
\newcommand{\imp}{\ensuremath{\rightarrow}\xspace}

\renewcommand{\iff}{\ensuremath{\leftrightarrow}\xspace}
\newcommand{\defas}{\ensuremath{\stackrel{\text{\tiny def}}{=}}\xspace}

\newcommand{\pos}{\phantom{\neg}}

\newcommand\mysout{\bgroup \markoverwith{{-}}\ULon}
\newcommand\nosout{\bgroup \markoverwith{{ }}\ULon}
\definecolor{mygray}{rgb}{0.90,0.90,0.90}
\definecolor{mywhite}{rgb}{1.00,1.00,1.00}

\renewcommand{\RSTODO}[1]{\noindent{\textcolor{blue}{{\fbox{RS TODO:} #1}}}}

\renewcommand{\TODO}[1]{\noindent{\textcolor{darkviolet}{{\fbox{TODO:} #1}}}}

\newcommand{\allx}{\ensuremath{\mathbf{x}}\xspace}

\newcommand{\ti}[1]{\ensuremath{\sf{t}^{(#1)}}\xspace}
\newcommand{\tn}[1]{\ti{n}}

\renewcommand{\TODO}[1]{\todo[inline,color=green!40]{{\small{#1}}}}

\renewcommand{\RSTODO}[1]{\todo[inline,color=green!40]{{\small{RS TODO: #1}}}}

\newcommand{\JDSIDENOTE}[1]{\todo[color=orange!40,size=tiny]{JD: #1}}

\newcommand{\myf}[1]{f_{#1}}

\renewcommand{\myf}[1]
{\ifthenelse%
 {\equal{#1}{11}}{x_1^2x_2}%
 {\ifthenelse%
  {\equal{#1}{12}}{x_1^3x_2}
  {\ifthenelse%
   {\equal{#1}{21}}{x_1x_2^2}
   {\ifthenelse%
    {\equal{#1}{22}}{x_1x_2^3}
    {\ifthenelse%
     {\equal{#1}{3}}{2x_1x_2}
     {3x_1x_2}
}}}}}

\newcommand{\OptFN}[2]{{\ifx&#2&\ensuremath{#1}\else\ensuremath{#1(#2)}\fi}}



%% file: rgb.tex
\definecolor {snow}                {rgb}{1.00,0.98,0.98}
\definecolor {ghostwhite}          {rgb}{0.97,0.97,1.00}
\definecolor {whitesmoke}          {rgb}{0.96,0.96,0.96}
\definecolor {gainsboro}           {rgb}{0.86,0.86,0.86}
\definecolor {floralwhite}         {rgb}{1.00,0.98,0.94}
\definecolor {oldlace}             {rgb}{0.99,0.96,0.90}
\definecolor {linen}               {rgb}{0.98,0.94,0.90}
\definecolor {antiquewhite}        {rgb}{0.98,0.92,0.84}
\definecolor {papayawhip}          {rgb}{1.00,0.94,0.84}
\definecolor {blanchedalmond}      {rgb}{1.00,0.92,0.80}
\definecolor {bisque}              {rgb}{1.00,0.89,0.77}
\definecolor {peachpuff}           {rgb}{1.00,0.85,0.73}
\definecolor {navajowhite}         {rgb}{1.00,0.87,0.68}
\definecolor {moccasin}            {rgb}{1.00,0.89,0.71}
\definecolor {cornsilk}            {rgb}{1.00,0.97,0.86}
\definecolor {ivory}               {rgb}{1.00,1.00,0.94}
\definecolor {lemonchiffon}        {rgb}{1.00,0.98,0.80}
\definecolor {seashell}            {rgb}{1.00,0.96,0.93}
\definecolor {honeydew}            {rgb}{0.94,1.00,0.94}
\definecolor {mintcream}           {rgb}{0.96,1.00,0.98}
\definecolor {azure}               {rgb}{0.94,1.00,1.00}
\definecolor {aliceblue}           {rgb}{0.94,0.97,1.00}
\definecolor {lavender}            {rgb}{0.90,0.90,0.98}
\definecolor {lavenderblush}       {rgb}{1.00,0.94,0.96}
\definecolor {mistyrose}           {rgb}{1.00,0.89,0.88}
\definecolor {white}               {rgb}{1.00,1.00,1.00}
\definecolor {black}               {rgb}{0.00,0.00,0.00}
\definecolor {darkslategray}       {rgb}{0.18,0.31,0.31}
\definecolor {dimgray}             {rgb}{0.41,0.41,0.41}
\definecolor {slategray}           {rgb}{0.44,0.50,0.56}
\definecolor {lightslategray}      {rgb}{0.47,0.53,0.60}
\definecolor {gray}                {rgb}{0.75,0.75,0.75}
\definecolor {lightgrey}           {rgb}{0.83,0.83,0.83}
\definecolor {midnightblue}        {rgb}{0.10,0.10,0.44}
\definecolor {navy}                {rgb}{0.00,0.00,0.50}
\definecolor {cornflowerblue}      {rgb}{0.39,0.58,0.93}
\definecolor {darkslateblue}       {rgb}{0.28,0.24,0.55}
\definecolor {slateblue}           {rgb}{0.42,0.35,0.80}
\definecolor {mediumslateblue}     {rgb}{0.48,0.41,0.93}
\definecolor {lightslateblue}      {rgb}{0.52,0.44,1.00}
\definecolor {mediumblue}          {rgb}{0.00,0.00,0.80}
\definecolor {royalblue}           {rgb}{0.25,0.41,0.88}
\definecolor {blue}                {rgb}{0.00,0.00,1.00}
\definecolor {dodgerblue}          {rgb}{0.12,0.56,1.00}
\definecolor {deepskyblue}         {rgb}{0.00,0.75,1.00}
\definecolor {skyblue}             {rgb}{0.53,0.81,0.92}
\definecolor {lightskyblue}        {rgb}{0.53,0.81,0.98}
\definecolor {steelblue}           {rgb}{0.27,0.51,0.71}
\definecolor {lightsteelblue}      {rgb}{0.69,0.77,0.87}
\definecolor {lightblue}           {rgb}{0.68,0.85,0.90}
\definecolor {powderblue}          {rgb}{0.69,0.88,0.90}
\definecolor {paleturquoise}       {rgb}{0.69,0.93,0.93}
\definecolor {darkturquoise}       {rgb}{0.00,0.81,0.82}
\definecolor {mediumturquoise}     {rgb}{0.28,0.82,0.80}
\definecolor {turquoise}           {rgb}{0.25,0.88,0.82}
\definecolor {cyan}                {rgb}{0.00,1.00,1.00}
\definecolor {lightcyan}           {rgb}{0.88,1.00,1.00}
\definecolor {cadetblue}           {rgb}{0.37,0.62,0.63}
\definecolor {mediumaquamarine}    {rgb}{0.40,0.80,0.67}
\definecolor {aquamarine}          {rgb}{0.50,1.00,0.83}
\definecolor {darkgreen}           {rgb}{0.00,0.39,0.00}
\definecolor {darkolivegreen}      {rgb}{0.33,0.42,0.18}
\definecolor {darkseagreen}        {rgb}{0.56,0.74,0.56}
\definecolor {seagreen}            {rgb}{0.18,0.55,0.34}
\definecolor {mediumseagreen}      {rgb}{0.24,0.70,0.44}
\definecolor {lightseagreen}       {rgb}{0.13,0.70,0.67}
\definecolor {palegreen}           {rgb}{0.60,0.98,0.60}
\definecolor {springgreen}         {rgb}{0.00,1.00,0.50}
\definecolor {lawngreen}           {rgb}{0.49,0.99,0.00}
\definecolor {green}               {rgb}{0.00,1.00,0.00}
\definecolor {chartreuse}          {rgb}{0.50,1.00,0.00}
\definecolor {mediumspringgreen}   {rgb}{0.00,0.98,0.60}
\definecolor {greenyellow}         {rgb}{0.68,1.00,0.18}
\definecolor {limegreen}           {rgb}{0.20,0.80,0.20}
\definecolor {yellowgreen}         {rgb}{0.60,0.80,0.20}
\definecolor {forestgreen}         {rgb}{0.13,0.55,0.13}
\definecolor {olivedrab}           {rgb}{0.42,0.56,0.14}
\definecolor {darkkhaki}           {rgb}{0.74,0.72,0.42}
\definecolor {khaki}               {rgb}{0.94,0.90,0.55}
\definecolor {palegoldenrod}       {rgb}{0.93,0.91,0.67}
\definecolor {lightgoldenrodyellow} {rgb}{0.98,0.98,0.82}
\definecolor {lightyellow}         {rgb}{1.00,1.00,0.88}
\definecolor {yellow}              {rgb}{1.00,1.00,0.00}
\definecolor {gold}                {rgb}{1.00,0.84,0.00}
\definecolor {lightgoldenrod}      {rgb}{0.93,0.87,0.51}
\definecolor {goldenrod}           {rgb}{0.85,0.65,0.13}
\definecolor {darkgoldenrod}       {rgb}{0.72,0.53,0.04}
\definecolor {rosybrown}           {rgb}{0.74,0.56,0.56}
\definecolor {indianred}           {rgb}{0.80,0.36,0.36}
\definecolor {saddlebrown}         {rgb}{0.55,0.27,0.07}
\definecolor {sienna}              {rgb}{0.63,0.32,0.18}
\definecolor {peru}                {rgb}{0.80,0.52,0.25}
\definecolor {burlywood}           {rgb}{0.87,0.72,0.53}
\definecolor {beige}               {rgb}{0.96,0.96,0.86}
\definecolor {wheat}               {rgb}{0.96,0.87,0.70}
\definecolor {sandybrown}          {rgb}{0.96,0.64,0.38}
\definecolor {tan}                 {rgb}{0.82,0.71,0.55}
\definecolor {chocolate}           {rgb}{0.82,0.41,0.12}
\definecolor {firebrick}           {rgb}{0.70,0.13,0.13}
\definecolor {brown}               {rgb}{0.65,0.16,0.16}
\definecolor {darksalmon}          {rgb}{0.91,0.59,0.48}
\definecolor {salmon}              {rgb}{0.98,0.50,0.45}
\definecolor {lightsalmon}         {rgb}{1.00,0.63,0.48}
\definecolor {orange}              {rgb}{1.00,0.65,0.00}
\definecolor {darkorange}          {rgb}{1.00,0.55,0.00}
\definecolor {coral}               {rgb}{1.00,0.50,0.31}
\definecolor {lightcoral}          {rgb}{0.94,0.50,0.50}
\definecolor {tomato}              {rgb}{1.00,0.39,0.28}
\definecolor {orangered}           {rgb}{1.00,0.27,0.00}
\definecolor {red}                 {rgb}{1.00,0.00,0.00}
\definecolor {hotpink}             {rgb}{1.00,0.41,0.71}
\definecolor {deeppink}            {rgb}{1.00,0.08,0.58}
\definecolor {pink}                {rgb}{1.00,0.75,0.80}
\definecolor {lightpink}           {rgb}{1.00,0.71,0.76}
\definecolor {palevioletred}       {rgb}{0.86,0.44,0.58}
\definecolor {maroon}              {rgb}{0.69,0.19,0.38}
\definecolor {mediumvioletred}     {rgb}{0.78,0.08,0.52}
\definecolor {violetred}           {rgb}{0.82,0.13,0.56}
\definecolor {magenta}             {rgb}{1.00,0.00,1.00}
\definecolor {violet}              {rgb}{0.93,0.51,0.93}
\definecolor {plum}                {rgb}{0.87,0.63,0.87}
\definecolor {orchid}              {rgb}{0.85,0.44,0.84}
\definecolor {mediumorchid}        {rgb}{0.73,0.33,0.83}
\definecolor {darkorchid}          {rgb}{0.60,0.20,0.80}
\definecolor {darkviolet}          {rgb}{0.58,0.00,0.83}
\definecolor {blueviolet}          {rgb}{0.54,0.17,0.89}
\definecolor {purple}              {rgb}{0.63,0.13,0.94}
\definecolor {mediumpurple}        {rgb}{0.58,0.44,0.86}
\definecolor {thistle}             {rgb}{0.85,0.75,0.85}
\definecolor {snow2}               {rgb}{0.93,0.91,0.91}
\definecolor {snow3}               {rgb}{0.80,0.79,0.79}
\definecolor {snow4}               {rgb}{0.55,0.54,0.54}
\definecolor {seashell2}           {rgb}{0.93,0.90,0.87}
\definecolor {seashell3}           {rgb}{0.80,0.77,0.75}
\definecolor {seashell4}           {rgb}{0.55,0.53,0.51}
\definecolor {antiquewhite1}       {rgb}{1.00,0.94,0.86}
\definecolor {antiquewhite2}       {rgb}{0.93,0.87,0.80}
\definecolor {antiquewhite3}       {rgb}{0.80,0.75,0.69}
\definecolor {antiquewhite4}       {rgb}{0.55,0.51,0.47}
\definecolor {bisque2}             {rgb}{0.93,0.84,0.72}
\definecolor {bisque3}             {rgb}{0.80,0.72,0.62}
\definecolor {bisque4}             {rgb}{0.55,0.49,0.42}
\definecolor {peachpuff2}          {rgb}{0.93,0.80,0.68}
\definecolor {peachpuff3}          {rgb}{0.80,0.69,0.58}
\definecolor {peachpuff4}          {rgb}{0.55,0.47,0.40}
\definecolor {navajowhite2}        {rgb}{0.93,0.81,0.63}
\definecolor {navajowhite3}        {rgb}{0.80,0.70,0.55}
\definecolor {navajowhite4}        {rgb}{0.55,0.47,0.37}
\definecolor {lemonchiffon2}       {rgb}{0.93,0.91,0.75}
\definecolor {lemonchiffon3}       {rgb}{0.80,0.79,0.65}
\definecolor {lemonchiffon4}       {rgb}{0.55,0.54,0.44}
\definecolor {cornsilk2}           {rgb}{0.93,0.91,0.80}
\definecolor {cornsilk3}           {rgb}{0.80,0.78,0.69}
\definecolor {cornsilk4}           {rgb}{0.55,0.53,0.47}
\definecolor {ivory2}              {rgb}{0.93,0.93,0.88}
\definecolor {ivory3}              {rgb}{0.80,0.80,0.76}
\definecolor {ivory4}              {rgb}{0.55,0.55,0.51}
\definecolor {honeydew2}           {rgb}{0.88,0.93,0.88}
\definecolor {honeydew3}           {rgb}{0.76,0.80,0.76}
\definecolor {honeydew4}           {rgb}{0.51,0.55,0.51}
\definecolor {lavenderblush2}      {rgb}{0.93,0.88,0.90}
\definecolor {lavenderblush3}      {rgb}{0.80,0.76,0.77}
\definecolor {lavenderblush4}      {rgb}{0.55,0.51,0.53}
\definecolor {mistyrose2}          {rgb}{0.93,0.84,0.82}
\definecolor {mistyrose3}          {rgb}{0.80,0.72,0.71}
\definecolor {mistyrose4}          {rgb}{0.55,0.49,0.48}
\definecolor {azure2}              {rgb}{0.88,0.93,0.93}
\definecolor {azure3}              {rgb}{0.76,0.80,0.80}
\definecolor {azure4}              {rgb}{0.51,0.55,0.55}
\definecolor {slateblue1}          {rgb}{0.51,0.44,1.00}
\definecolor {slateblue2}          {rgb}{0.48,0.40,0.93}
\definecolor {slateblue3}          {rgb}{0.41,0.35,0.80}
\definecolor {slateblue4}          {rgb}{0.28,0.24,0.55}
\definecolor {royalblue1}          {rgb}{0.28,0.46,1.00}
\definecolor {royalblue2}          {rgb}{0.26,0.43,0.93}
\definecolor {royalblue3}          {rgb}{0.23,0.37,0.80}
\definecolor {royalblue4}          {rgb}{0.15,0.25,0.55}
\definecolor {blue2}               {rgb}{0.00,0.00,0.93}
\definecolor {blue4}               {rgb}{0.00,0.00,0.55}
\definecolor {dodgerblue2}         {rgb}{0.11,0.53,0.93}
\definecolor {dodgerblue3}         {rgb}{0.09,0.45,0.80}
\definecolor {dodgerblue4}         {rgb}{0.06,0.31,0.55}
\definecolor {steelblue1}          {rgb}{0.39,0.72,1.00}
\definecolor {steelblue2}          {rgb}{0.36,0.67,0.93}
\definecolor {steelblue3}          {rgb}{0.31,0.58,0.80}
\definecolor {steelblue4}          {rgb}{0.21,0.39,0.55}
\definecolor {deepskyblue2}        {rgb}{0.00,0.70,0.93}
\definecolor {deepskyblue3}        {rgb}{0.00,0.60,0.80}
\definecolor {deepskyblue4}        {rgb}{0.00,0.41,0.55}
\definecolor {skyblue1}            {rgb}{0.53,0.81,1.00}
\definecolor {skyblue2}            {rgb}{0.49,0.75,0.93}
\definecolor {skyblue3}            {rgb}{0.42,0.65,0.80}
\definecolor {skyblue4}            {rgb}{0.29,0.44,0.55}
\definecolor {lightskyblue1}       {rgb}{0.69,0.89,1.00}
\definecolor {lightskyblue2}       {rgb}{0.64,0.83,0.93}
\definecolor {lightskyblue3}       {rgb}{0.55,0.71,0.80}
\definecolor {lightskyblue4}       {rgb}{0.38,0.48,0.55}
\definecolor {slategray1}          {rgb}{0.78,0.89,1.00}
\definecolor {slategray2}          {rgb}{0.73,0.83,0.93}
\definecolor {slategray3}          {rgb}{0.62,0.71,0.80}
\definecolor {slategray4}          {rgb}{0.42,0.48,0.55}
\definecolor {lightsteelblue1}     {rgb}{0.79,0.88,1.00}
\definecolor {lightsteelblue2}     {rgb}{0.74,0.82,0.93}
\definecolor {lightsteelblue3}     {rgb}{0.64,0.71,0.80}
\definecolor {lightsteelblue4}     {rgb}{0.43,0.48,0.55}
\definecolor {lightblue1}          {rgb}{0.75,0.94,1.00}
\definecolor {lightblue2}          {rgb}{0.70,0.87,0.93}
\definecolor {lightblue3}          {rgb}{0.60,0.75,0.80}
\definecolor {lightblue4}          {rgb}{0.41,0.51,0.55}
\definecolor {lightcyan2}          {rgb}{0.82,0.93,0.93}
\definecolor {lightcyan3}          {rgb}{0.71,0.80,0.80}
\definecolor {lightcyan4}          {rgb}{0.48,0.55,0.55}
\definecolor {paleturquoise1}      {rgb}{0.73,1.00,1.00}
\definecolor {paleturquoise2}      {rgb}{0.68,0.93,0.93}
\definecolor {paleturquoise3}      {rgb}{0.59,0.80,0.80}
\definecolor {paleturquoise4}      {rgb}{0.40,0.55,0.55}
\definecolor {cadetblue1}          {rgb}{0.60,0.96,1.00}
\definecolor {cadetblue2}          {rgb}{0.56,0.90,0.93}
\definecolor {cadetblue3}          {rgb}{0.48,0.77,0.80}
\definecolor {cadetblue4}          {rgb}{0.33,0.53,0.55}
\definecolor {turquoise1}          {rgb}{0.00,0.96,1.00}
\definecolor {turquoise2}          {rgb}{0.00,0.90,0.93}
\definecolor {turquoise3}          {rgb}{0.00,0.77,0.80}
\definecolor {turquoise4}          {rgb}{0.00,0.53,0.55}
\definecolor {cyan2}               {rgb}{0.00,0.93,0.93}
\definecolor {cyan3}               {rgb}{0.00,0.80,0.80}
\definecolor {cyan4}               {rgb}{0.00,0.55,0.55}
\definecolor {darkslategray1}      {rgb}{0.59,1.00,1.00}
\definecolor {darkslategray2}      {rgb}{0.55,0.93,0.93}
\definecolor {darkslategray3}      {rgb}{0.47,0.80,0.80}
\definecolor {darkslategray4}      {rgb}{0.32,0.55,0.55}
\definecolor {aquamarine2}         {rgb}{0.46,0.93,0.78}
\definecolor {aquamarine4}         {rgb}{0.27,0.55,0.45}
\definecolor {darkseagreen1}       {rgb}{0.76,1.00,0.76}
\definecolor {darkseagreen2}       {rgb}{0.71,0.93,0.71}
\definecolor {darkseagreen3}       {rgb}{0.61,0.80,0.61}
\definecolor {darkseagreen4}       {rgb}{0.41,0.55,0.41}
\definecolor {seagreen1}           {rgb}{0.33,1.00,0.62}
\definecolor {seagreen2}           {rgb}{0.31,0.93,0.58}
\definecolor {seagreen3}           {rgb}{0.26,0.80,0.50}
\definecolor {palegreen1}          {rgb}{0.60,1.00,0.60}
\definecolor {palegreen2}          {rgb}{0.56,0.93,0.56}
\definecolor {palegreen3}          {rgb}{0.49,0.80,0.49}
\definecolor {palegreen4}          {rgb}{0.33,0.55,0.33}
\definecolor {springgreen2}        {rgb}{0.00,0.93,0.46}
\definecolor {springgreen3}        {rgb}{0.00,0.80,0.40}
\definecolor {springgreen4}        {rgb}{0.00,0.55,0.27}
\definecolor {green2}              {rgb}{0.00,0.93,0.00}
\definecolor {green3}              {rgb}{0.00,0.80,0.00}
\definecolor {green4}              {rgb}{0.00,0.55,0.00}
\definecolor {chartreuse2}         {rgb}{0.46,0.93,0.00}
\definecolor {chartreuse3}         {rgb}{0.40,0.80,0.00}
\definecolor {chartreuse4}         {rgb}{0.27,0.55,0.00}
\definecolor {olivedrab1}          {rgb}{0.75,1.00,0.24}
\definecolor {olivedrab2}          {rgb}{0.70,0.93,0.23}
\definecolor {olivedrab4}          {rgb}{0.41,0.55,0.13}
\definecolor {darkolivegreen1}     {rgb}{0.79,1.00,0.44}
\definecolor {darkolivegreen2}     {rgb}{0.74,0.93,0.41}
\definecolor {darkolivegreen3}     {rgb}{0.64,0.80,0.35}
\definecolor {darkolivegreen4}     {rgb}{0.43,0.55,0.24}
\definecolor {khaki1}              {rgb}{1.00,0.96,0.56}
\definecolor {khaki2}              {rgb}{0.93,0.90,0.52}
\definecolor {khaki3}              {rgb}{0.80,0.78,0.45}
\definecolor {khaki4}              {rgb}{0.55,0.53,0.31}
\definecolor {lightgoldenrod1}     {rgb}{1.00,0.93,0.55}
\definecolor {lightgoldenrod2}     {rgb}{0.93,0.86,0.51}
\definecolor {lightgoldenrod3}     {rgb}{0.80,0.75,0.44}
\definecolor {lightgoldenrod4}     {rgb}{0.55,0.51,0.30}
\definecolor {lightyellow2}        {rgb}{0.93,0.93,0.82}
\definecolor {lightyellow3}        {rgb}{0.80,0.80,0.71}
\definecolor {lightyellow4}        {rgb}{0.55,0.55,0.48}
\definecolor {yellow2}             {rgb}{0.93,0.93,0.00}
\definecolor {yellow3}             {rgb}{0.80,0.80,0.00}
\definecolor {yellow4}             {rgb}{0.55,0.55,0.00}
\definecolor {gold2}               {rgb}{0.93,0.79,0.00}
\definecolor {gold3}               {rgb}{0.80,0.68,0.00}
\definecolor {gold4}               {rgb}{0.55,0.46,0.00}
\definecolor {goldenrod1}          {rgb}{1.00,0.76,0.15}
\definecolor {goldenrod2}          {rgb}{0.93,0.71,0.13}
\definecolor {goldenrod3}          {rgb}{0.80,0.61,0.11}
\definecolor {goldenrod4}          {rgb}{0.55,0.41,0.08}
\definecolor {darkgoldenrod1}      {rgb}{1.00,0.73,0.06}
\definecolor {darkgoldenrod2}      {rgb}{0.93,0.68,0.05}
\definecolor {darkgoldenrod3}      {rgb}{0.80,0.58,0.05}
\definecolor {darkgoldenrod4}      {rgb}{0.55,0.40,0.03}
\definecolor {rosybrown1}          {rgb}{1.00,0.76,0.76}
\definecolor {rosybrown2}          {rgb}{0.93,0.71,0.71}
\definecolor {rosybrown3}          {rgb}{0.80,0.61,0.61}
\definecolor {rosybrown4}          {rgb}{0.55,0.41,0.41}
\definecolor {indianred1}          {rgb}{1.00,0.42,0.42}
\definecolor {indianred2}          {rgb}{0.93,0.39,0.39}
\definecolor {indianred3}          {rgb}{0.80,0.33,0.33}
\definecolor {indianred4}          {rgb}{0.55,0.23,0.23}
\definecolor {sienna1}             {rgb}{1.00,0.51,0.28}
\definecolor {sienna2}             {rgb}{0.93,0.47,0.26}
\definecolor {sienna3}             {rgb}{0.80,0.41,0.22}
\definecolor {sienna4}             {rgb}{0.55,0.28,0.15}
\definecolor {burlywood1}          {rgb}{1.00,0.83,0.61}
\definecolor {burlywood2}          {rgb}{0.93,0.77,0.57}
\definecolor {burlywood3}          {rgb}{0.80,0.67,0.49}
\definecolor {burlywood4}          {rgb}{0.55,0.45,0.33}
\definecolor {wheat1}              {rgb}{1.00,0.91,0.73}
\definecolor {wheat2}              {rgb}{0.93,0.85,0.68}
\definecolor {wheat3}              {rgb}{0.80,0.73,0.59}
\definecolor {wheat4}              {rgb}{0.55,0.49,0.40}
\definecolor {tan1}                {rgb}{1.00,0.65,0.31}
\definecolor {tan2}                {rgb}{0.93,0.60,0.29}
\definecolor {tan4}                {rgb}{0.55,0.35,0.17}
\definecolor {chocolate1}          {rgb}{1.00,0.50,0.14}
\definecolor {chocolate2}          {rgb}{0.93,0.46,0.13}
\definecolor {chocolate3}          {rgb}{0.80,0.40,0.11}
\definecolor {firebrick1}          {rgb}{1.00,0.19,0.19}
\definecolor {firebrick2}          {rgb}{0.93,0.17,0.17}
\definecolor {firebrick3}          {rgb}{0.80,0.15,0.15}
\definecolor {firebrick4}          {rgb}{0.55,0.10,0.10}
\definecolor {brown1}              {rgb}{1.00,0.25,0.25}
\definecolor {brown2}              {rgb}{0.93,0.23,0.23}
\definecolor {brown3}              {rgb}{0.80,0.20,0.20}
\definecolor {brown4}              {rgb}{0.55,0.14,0.14}
\definecolor {salmon1}             {rgb}{1.00,0.55,0.41}
\definecolor {salmon2}             {rgb}{0.93,0.51,0.38}
\definecolor {salmon3}             {rgb}{0.80,0.44,0.33}
\definecolor {salmon4}             {rgb}{0.55,0.30,0.22}
\definecolor {lightsalmon2}        {rgb}{0.93,0.58,0.45}
\definecolor {lightsalmon3}        {rgb}{0.80,0.51,0.38}
\definecolor {lightsalmon4}        {rgb}{0.55,0.34,0.26}
\definecolor {orange2}             {rgb}{0.93,0.60,0.00}
\definecolor {orange3}             {rgb}{0.80,0.52,0.00}
\definecolor {orange4}             {rgb}{0.55,0.35,0.00}
\definecolor {darkorange1}         {rgb}{1.00,0.50,0.00}
\definecolor {darkorange2}         {rgb}{0.93,0.46,0.00}
\definecolor {darkorange3}         {rgb}{0.80,0.40,0.00}
\definecolor {darkorange4}         {rgb}{0.55,0.27,0.00}
\definecolor {coral1}              {rgb}{1.00,0.45,0.34}
\definecolor {coral2}              {rgb}{0.93,0.42,0.31}
\definecolor {coral3}              {rgb}{0.80,0.36,0.27}
\definecolor {coral4}              {rgb}{0.55,0.24,0.18}
\definecolor {tomato2}             {rgb}{0.93,0.36,0.26}
\definecolor {tomato3}             {rgb}{0.80,0.31,0.22}
\definecolor {tomato4}             {rgb}{0.55,0.21,0.15}
\definecolor {orangered2}          {rgb}{0.93,0.25,0.00}
\definecolor {orangered3}          {rgb}{0.80,0.22,0.00}
\definecolor {orangered4}          {rgb}{0.55,0.15,0.00}
\definecolor {red2}                {rgb}{0.93,0.00,0.00}
\definecolor {red3}                {rgb}{0.80,0.00,0.00}
\definecolor {red4}                {rgb}{0.55,0.00,0.00}
\definecolor {deeppink2}           {rgb}{0.93,0.07,0.54}
\definecolor {deeppink3}           {rgb}{0.80,0.06,0.46}
\definecolor {deeppink4}           {rgb}{0.55,0.04,0.31}
\definecolor {hotpink1}            {rgb}{1.00,0.43,0.71}
\definecolor {hotpink2}            {rgb}{0.93,0.42,0.65}
\definecolor {hotpink3}            {rgb}{0.80,0.38,0.56}
\definecolor {hotpink4}            {rgb}{0.55,0.23,0.38}
\definecolor {pink1}               {rgb}{1.00,0.71,0.77}
\definecolor {pink2}               {rgb}{0.93,0.66,0.72}
\definecolor {pink3}               {rgb}{0.80,0.57,0.62}
\definecolor {pink4}               {rgb}{0.55,0.39,0.42}
\definecolor {lightpink1}          {rgb}{1.00,0.68,0.73}
\definecolor {lightpink2}          {rgb}{0.93,0.64,0.68}
\definecolor {lightpink3}          {rgb}{0.80,0.55,0.58}
\definecolor {lightpink4}          {rgb}{0.55,0.37,0.40}
\definecolor {palevioletred1}      {rgb}{1.00,0.51,0.67}
\definecolor {palevioletred2}      {rgb}{0.93,0.47,0.62}
\definecolor {palevioletred3}      {rgb}{0.80,0.41,0.54}
\definecolor {palevioletred4}      {rgb}{0.55,0.28,0.36}
\definecolor {maroon1}             {rgb}{1.00,0.20,0.70}
\definecolor {maroon2}             {rgb}{0.93,0.19,0.65}
\definecolor {maroon3}             {rgb}{0.80,0.16,0.56}
\definecolor {maroon4}             {rgb}{0.55,0.11,0.38}
\definecolor {violetred1}          {rgb}{1.00,0.24,0.59}
\definecolor {violetred2}          {rgb}{0.93,0.23,0.55}
\definecolor {violetred3}          {rgb}{0.80,0.20,0.47}
\definecolor {violetred4}          {rgb}{0.55,0.13,0.32}
\definecolor {magenta2}            {rgb}{0.93,0.00,0.93}
\definecolor {magenta3}            {rgb}{0.80,0.00,0.80}
\definecolor {magenta4}            {rgb}{0.55,0.00,0.55}
\definecolor {orchid1}             {rgb}{1.00,0.51,0.98}
\definecolor {orchid2}             {rgb}{0.93,0.48,0.91}
\definecolor {orchid3}             {rgb}{0.80,0.41,0.79}
\definecolor {orchid4}             {rgb}{0.55,0.28,0.54}
\definecolor {plum1}               {rgb}{1.00,0.73,1.00}
\definecolor {plum2}               {rgb}{0.93,0.68,0.93}
\definecolor {plum3}               {rgb}{0.80,0.59,0.80}
\definecolor {plum4}               {rgb}{0.55,0.40,0.55}
\definecolor {mediumorchid1}       {rgb}{0.88,0.40,1.00}
\definecolor {mediumorchid2}       {rgb}{0.82,0.37,0.93}
\definecolor {mediumorchid3}       {rgb}{0.71,0.32,0.80}
\definecolor {mediumorchid4}       {rgb}{0.48,0.22,0.55}
\definecolor {darkorchid1}         {rgb}{0.75,0.24,1.00}
\definecolor {darkorchid2}         {rgb}{0.70,0.23,0.93}
\definecolor {darkorchid3}         {rgb}{0.60,0.20,0.80}
\definecolor {darkorchid4}         {rgb}{0.41,0.13,0.55}
\definecolor {purple1}             {rgb}{0.61,0.19,1.00}
\definecolor {purple2}             {rgb}{0.57,0.17,0.93}
\definecolor {purple3}             {rgb}{0.49,0.15,0.80}
\definecolor {purple4}             {rgb}{0.33,0.10,0.55}
\definecolor {mediumpurple1}       {rgb}{0.67,0.51,1.00}
\definecolor {mediumpurple2}       {rgb}{0.62,0.47,0.93}
\definecolor {mediumpurple3}       {rgb}{0.54,0.41,0.80}
\definecolor {mediumpurple4}       {rgb}{0.36,0.28,0.55}
\definecolor {thistle1}            {rgb}{1.00,0.88,1.00}
\definecolor {thistle2}            {rgb}{0.93,0.82,0.93}
\definecolor {thistle3}            {rgb}{0.80,0.71,0.80}
\definecolor {thistle4}            {rgb}{0.55,0.48,0.55}
\definecolor {gray1}               {rgb}{0.01,0.01,0.01}
\definecolor {gray2}               {rgb}{0.02,0.02,0.02}
\definecolor {gray3}               {rgb}{0.03,0.03,0.03}
\definecolor {gray4}               {rgb}{0.04,0.04,0.04}
\definecolor {gray5}               {rgb}{0.05,0.05,0.05}
\definecolor {gray6}               {rgb}{0.06,0.06,0.06}
\definecolor {gray7}               {rgb}{0.07,0.07,0.07}
\definecolor {gray8}               {rgb}{0.08,0.08,0.08}
\definecolor {gray9}               {rgb}{0.09,0.09,0.09}
\definecolor {gray10}              {rgb}{0.10,0.10,0.10}
\definecolor {gray11}              {rgb}{0.11,0.11,0.11}
\definecolor {gray12}              {rgb}{0.12,0.12,0.12}
\definecolor {gray13}              {rgb}{0.13,0.13,0.13}
\definecolor {gray14}              {rgb}{0.14,0.14,0.14}
\definecolor {gray15}              {rgb}{0.15,0.15,0.15}
\definecolor {gray16}              {rgb}{0.16,0.16,0.16}
\definecolor {gray17}              {rgb}{0.17,0.17,0.17}
\definecolor {gray18}              {rgb}{0.18,0.18,0.18}
\definecolor {gray19}              {rgb}{0.19,0.19,0.19}
\definecolor {gray20}              {rgb}{0.20,0.20,0.20}
\definecolor {gray21}              {rgb}{0.21,0.21,0.21}
\definecolor {gray22}              {rgb}{0.22,0.22,0.22}
\definecolor {gray23}              {rgb}{0.23,0.23,0.23}
\definecolor {gray24}              {rgb}{0.24,0.24,0.24}
\definecolor {gray25}              {rgb}{0.25,0.25,0.25}
\definecolor {gray26}              {rgb}{0.26,0.26,0.26}
\definecolor {gray27}              {rgb}{0.27,0.27,0.27}
\definecolor {gray28}              {rgb}{0.28,0.28,0.28}
\definecolor {gray29}              {rgb}{0.29,0.29,0.29}
\definecolor {gray30}              {rgb}{0.30,0.30,0.30}
\definecolor {gray31}              {rgb}{0.31,0.31,0.31}
\definecolor {gray32}              {rgb}{0.32,0.32,0.32}
\definecolor {gray33}              {rgb}{0.33,0.33,0.33}
\definecolor {gray34}              {rgb}{0.34,0.34,0.34}
\definecolor {gray35}              {rgb}{0.35,0.35,0.35}
\definecolor {gray36}              {rgb}{0.36,0.36,0.36}
\definecolor {gray37}              {rgb}{0.37,0.37,0.37}
\definecolor {gray38}              {rgb}{0.38,0.38,0.38}
\definecolor {gray39}              {rgb}{0.39,0.39,0.39}
\definecolor {gray40}              {rgb}{0.40,0.40,0.40}
\definecolor {gray42}              {rgb}{0.42,0.42,0.42}
\definecolor {gray43}              {rgb}{0.43,0.43,0.43}
\definecolor {gray44}              {rgb}{0.44,0.44,0.44}
\definecolor {gray45}              {rgb}{0.45,0.45,0.45}
\definecolor {gray46}              {rgb}{0.46,0.46,0.46}
\definecolor {gray47}              {rgb}{0.47,0.47,0.47}
\definecolor {gray48}              {rgb}{0.48,0.48,0.48}
\definecolor {gray49}              {rgb}{0.49,0.49,0.49}
\definecolor {gray50}              {rgb}{0.50,0.50,0.50}
\definecolor {gray51}              {rgb}{0.51,0.51,0.51}
\definecolor {gray52}              {rgb}{0.52,0.52,0.52}
\definecolor {gray53}              {rgb}{0.53,0.53,0.53}
\definecolor {gray54}              {rgb}{0.54,0.54,0.54}
\definecolor {gray55}              {rgb}{0.55,0.55,0.55}
\definecolor {gray56}              {rgb}{0.56,0.56,0.56}
\definecolor {gray57}              {rgb}{0.57,0.57,0.57}
\definecolor {gray58}              {rgb}{0.58,0.58,0.58}
\definecolor {gray59}              {rgb}{0.59,0.59,0.59}
\definecolor {gray60}              {rgb}{0.60,0.60,0.60}
\definecolor {gray61}              {rgb}{0.61,0.61,0.61}
\definecolor {gray62}              {rgb}{0.62,0.62,0.62}
\definecolor {gray63}              {rgb}{0.63,0.63,0.63}
\definecolor {gray64}              {rgb}{0.64,0.64,0.64}
\definecolor {gray65}              {rgb}{0.65,0.65,0.65}
\definecolor {gray66}              {rgb}{0.66,0.66,0.66}
\definecolor {gray67}              {rgb}{0.67,0.67,0.67}
\definecolor {gray68}              {rgb}{0.68,0.68,0.68}
\definecolor {gray69}              {rgb}{0.69,0.69,0.69}
\definecolor {gray70}              {rgb}{0.70,0.70,0.70}
\definecolor {gray71}              {rgb}{0.71,0.71,0.71}
\definecolor {gray72}              {rgb}{0.72,0.72,0.72}
\definecolor {gray73}              {rgb}{0.73,0.73,0.73}
\definecolor {gray74}              {rgb}{0.74,0.74,0.74}
\definecolor {gray75}              {rgb}{0.75,0.75,0.75}
\definecolor {gray76}              {rgb}{0.76,0.76,0.76}
\definecolor {gray77}              {rgb}{0.77,0.77,0.77}
\definecolor {gray78}              {rgb}{0.78,0.78,0.78}
\definecolor {gray79}              {rgb}{0.79,0.79,0.79}
\definecolor {gray80}              {rgb}{0.80,0.80,0.80}
\definecolor {gray81}              {rgb}{0.81,0.81,0.81}
\definecolor {gray82}              {rgb}{0.82,0.82,0.82}
\definecolor {gray83}              {rgb}{0.83,0.83,0.83}
\definecolor {gray84}              {rgb}{0.84,0.84,0.84}
\definecolor {gray85}              {rgb}{0.85,0.85,0.85}
\definecolor {gray86}              {rgb}{0.86,0.86,0.86}
\definecolor {gray87}              {rgb}{0.87,0.87,0.87}
\definecolor {gray88}              {rgb}{0.88,0.88,0.88}
\definecolor {gray89}              {rgb}{0.89,0.89,0.89}
\definecolor {gray90}              {rgb}{0.90,0.90,0.90}
\definecolor {gray91}              {rgb}{0.91,0.91,0.91}
\definecolor {gray92}              {rgb}{0.92,0.92,0.92}
\definecolor {gray93}              {rgb}{0.93,0.93,0.93}
\definecolor {gray94}              {rgb}{0.94,0.94,0.94}
\definecolor {gray95}              {rgb}{0.95,0.95,0.95}
\definecolor {gray97}              {rgb}{0.97,0.97,0.97}
\definecolor {gray98}              {rgb}{0.98,0.98,0.98}
\definecolor {gray99}              {rgb}{0.99,0.99,0.99}
\definecolor {darkgrey}            {rgb}{0.66,0.66,0.66}

%% file: intro.tex
\section*{Introduction}

Integer factorization (IF) is the problem of factoring a positive integer into a product of small integers, called factors. If the factors are restricted to be prime,
we refer to it as prime factorization (PF).
Finding the prime factors of prime numbers becomes increasingly difficult as the numbers get larger. 
This difficulty is exploited in modern cryptography, where prime
factorization is used as a basis for secure encryption algorithms
(e.g. the RSA public-key encryption \cite{rivest1978method}) since the
process of factoring large numbers is currently considered
computationally infeasible for classical computers.

Quantum computers have the potential to perform PF exponentially faster than classical computers. A first approach in tackling PF by quantum computing is \emph{Shor's algorithm} \cite{shor94}. 
This technique takes advantage of the properties of quantum mechanics,
such as superposition and entanglement, to factor numbers into
their prime factors in poly-logarithmic time.
Although several efforts in implementing this algorithm, and variations
thereof,  on
  existing gate-based quantum computers have been presented in the
  literature~\cite{Vandersypen2001,Lucero2012,Martín-López2012,doi:10.1126/science.aad9480,PhysRevA.100.012305},
  plus other approaches\cite{selvarajan21},
the size of IP/PF which were actually implemented and solved on pure quantum
devices is very small, in the order of few thousands.
(Notice that a large-scale simulation of Shor's algorithm of GPU-based classical
supercomputer allowed to factorize up to 549,755,813,701
\cite{math11194222}, and that 
the factorization of the single number 1,099,551,473,989 was made possible  by means of
hybrid quantum-classical algorithms~\cite{Karamlou21}.)
%
%
{\em Quantum Annealing (QA)} has shown to be effective in performing
prime factorization, e.g., by reducing high-degree cost functions to quadratic either by using Groebner bases \cite{Dridi2017} or by using equivalent
quadratic models produced by adding ancillary variables
\cite{Jiang2018}, or by related approaches \cite{mengoni2020breaking}.
Currently, the largest factorization
problem mapped to the quantum annealer D-Wave 2000Q is 376,289. 
Moreover, all bi-primes up to 200,000 have been
solved by D-Wave 2X processors \cite{Dridi2017,Jiang2018}.
{Also, by using D-Wave hybrid Classic-QA tool, $1,005,973$ has been factored  \cite{Wang20}.}
\ignore{Notice that the state-of-the-art Advantage systems, D-Wave
latest-generation QAs, extend the older D-Wave 2X systems by providing more than 5000 qubits
and more complex connectivity patterns \cite{Boothby2019}.}
(See Willsch et al.\cite{math11194222} for a recent very-detailed survey on solving PF with quantum devices.)  

In this paper, we propose a novel approach based on a modular
version of locally-structured embedding
of satisfiability problems~\cite{Bian2016,Bian2020}  to encode
IF/PF problems into Ising models and
solve them using QA. Our contribution is twofold.  

First,  we present a novel modular {\em encoding} of a binary multiplier circuit into
the architecture of the most recent D-Wave QA devices. The key contribution is a compact encoding of a controlled full-adder
into an 8-qubit module in the Pegasus topology \cite{Boothby2019}, which we synthesized offline by means of
Optimization Modulo Theories. The multiplier circuit is then built by
exploiting a bunch of novel ideas, namely {\em alternating modules}, {\em qubit
sharing} between neighboring modules, and {\em virtual chaining}
between non-coupled qubits.
This allows us to encode up to a 21×12-bit multiplier (resp. a 22×8-bit one)
into the Pegasus 5760-qubit topology of current annealers, so that a faulty-free annealer could be
fed an integer factorization problem up to
$8,587,833,345 = 2,097,151 \times 4,095$ 
{(resp.  $1,069,547,265 = 4,194,303\times255$)), 
allowing for
prime factorization of up to 
{$8,583,606,299=2,097,143 \times 4,093$}
 (resp. $1,052,769,551 = 4,194,301\times251$).}
To the best of our knowledge, these are the largest
factorization problems ever encoded into a quantum annealer.
{We stress the fact that, given the modularity of the
  encoding, this number scales up automatically with the growth of the
qubit number in the chip.}

  Second, we have
  investigated the problem of actually {\em solving} encoded PF problems
  by running an extensive experimental evaluation on
  a D-Wave Advantage 4.1 quantum annealer.
  %
  Due to faulty qubits and
qubit couplings of the QA hardware we had access to, it was possible
to feed to it at most a 17×8-bit multiplier, corresponding to at most
a 33,423,105 = 131,071 $\times$ 255 factorization. In order to help
the annealer in reaching the global minimum, in the
experiments we introduced different approaches to initialize the
multiplier qubits 
and adopted several 
performance enhancement techniques, like {\em thermal relaxation},
{\em pausing}, and {\em reverse
annealing}, which we combined together by iterative strategies,
discussing their synergy when combined.  
Overall, exploiting all the encoding and solving techniques described
in this paper, 
{$8,219,999=32,749 \times 251$}
was the highest prime product we were
 able to factorize within the limits of our QPU resources.
 To the best of our knowledge, this is the largest number which
 was  ever factorized by means of a quantum annealer, and
   more generally by a quantum device, without adopting hybrid
 quantum-classical techniques.

\smallskip
\noindent {\em Disclaimer.} Due to space constraints, some details
  in some figures may not be easy to grasp from a printed
  version of this paper. Nevertheless, all figures are
  high-resolution ones, so that every detail can be grasped in full
if they are seen via a pdf viewer.

%% file: foundations.tex
\section*{Foundations}
\subsection*{D-Wave quantum annealers}

\JSChange{
{From a physicist's perspective, D-Wave's quantum annealers (QAs) are
quantum devices 
that use quantum phenomena
to reach minimum-energy states in terms of the values of their {\em qubits}
(i.e. minimum-energy states of superconductong loops).}
%
For a 
these QAs, the (quantum) {\it Hamiltonian} $H(s)$
}
---which corresponds to the classical Hamiltonian
that described some physical system in terms of its energies---
\JSChange{
is represented by the sum of 
the driver Hamiltonian $H_{driver}$
and the classical Ising Hamiltonian $H_{Ising}$,
where $\hat\sigma_{x,z}^{(i)}$ are Pauli matrices operating on a qubit
$q_i$, s.t. 
$h_i$ and $J_{i, j}$ are programmable parameters representing the qubit biases and coupling strengths:
\begin{eqnarray}\label{eq: 1}\textstyle
H(s) \defas -\frac{A(s)}{2} H_{driver} + \frac{B(s)}{2} H_{Ising},\ \ \ \ s.t.\ 
H_{driver} \defas \sum_i \hat\sigma_x^{(i)}, \
H_{Ising} \defas \sum_i h_i \hat\sigma_z^{(i)} + \sum_{i>j} J_{ij} \hat\sigma_z^{(i)} \hat\sigma_z^{(j)}.
\end{eqnarray}
The parameter $s$ is the normalized anneal fraction, $s=t/t_f \in [0, 1]$,
where $t$ is time and $t_f$ is the total time of the anneal process.
This $s$-dependent Hamiltonian $H(s)$
smoothly interpolates
between $H_{driver}$ and $H_{Ising}$
through the two annealing functions $A(s), B(s)$,
as shown in Figure \ref{fig: annealing-schedule}.
At $s=0$, the system starts 
in the groundstate of $H_{driver}$, 
with all qubits in the superposition state of 0 and 1;
as the system is annealed $s \uparrow$,
the dominance of $H_{driver}$ decreases
and $H_{Ising}$ comes to play;
at the end of the annealing process $s=1$,
the system would end up in 
{a classical state that corresponds to $H_{Ising}$.}
According to the quantum adiabatic theorem,
the system will remain in the instantaneous groundstate
through the evolution iff the system is annealed slowly enough.
The required runtime according to the theorem
is proportional to $\frac{1}{gap^2}$,
where $gap$ is the minimal gap
between the ground state and excited states
during the system evolution.
}

From a computer scientist's perspective, D-Wave's QAs are specialized quantum computers 
  which draw optima or
near-optima from quadratic cost functions on binary variables, that
is, specialized hardware for
solving the \emph{Ising {problem}}:\cite{Bian2020}
\begin{eqnarray}
          \label{eq:isingmodel_minimization}\textstyle
 argmin_{\zs \in \set{-1,1}^{|V|}}\ H(\zs), && \textstyle
\label{eq:isingmodel}
s.t. \ H(\zs) \defas  \sum_{i \in V} h_i z_i +
                 \sum_{\substack{(i,j)\in E}} J_{i,j} z_i z_j,
\end{eqnarray}
where each variable $z_i\in\set{-1,1}$ is associated with a qubit;
$G=\tuple{V,E}$ is an undirected graph, {\em the hardware graph} or
{\em topology}, whose
edges correspond to the physically-allowed qubit interactions; and
$ h_i$, $J_{i,j}$ are programmable real-valued parameters.
%
\ignore{
The
Chimera topology \JDSIDENOTE{Add reference} was first introduced in
the D-Wave One quantum annealing machine and is based on a
two-dimensional lattice of qubits. The lattice is divided into cells,
where each cell contains eight qubits arranged in a bipartite
graph. We call qubits on the same side of the partition
either vertical or horizontal. This
connectivity pattern results in a complex, sparse graph of qubits with
a hierarchical structure, allowing the system to perform large-scale
optimization problems with high connectivity and low error rates. The
ranges of values for biases and couplings are set respectively to
$[-2,2]$ and $[-1,1]$.
}
The current Pegasus topology \cite{Boothby2019} was introduced in the D-Wave
Advantage
quantum annealing machine and is based on a lattice of qubits.
The lattice is divided into cells (``tiles''),
where each cell contains eight qubits arranged in a bipartite
graph. We call qubits on the same side of the partition
either {\em vertical} or {\em horizontal} qubits. Qubits of the same side inside each tile
are connected 2-by-2. 
Figure \ref{fig: pegasus} shows the Pegasus topology for a $3\times 3$
subgraph.  It extends the previous Chimera topology by adding more
connections between the tiles so that the degree of connectivity of
each qubit is up to 15. In particular, each tile is now connected to
diagonally neighboring tiles through $45^\circ$,  $120^\circ$ and $150^\circ$
connections among qubits w.r.t. the $x$ axis (we
will refer to them as {\em diagonal couplings}). Moreover,
the configurable range of coefficients also increases, 
e.g., D-Wave Advantage 4.1 systems allow for biases and couplings
s.t. $h_i\in[-4, 4]$ and $J_{i,j}\in[-2, 1]$.

\begin{figure}
\centering
\subfloat[Annealing Schedule]{\label{fig: annealing-schedule}
\includegraphics[width=0.45\textwidth]{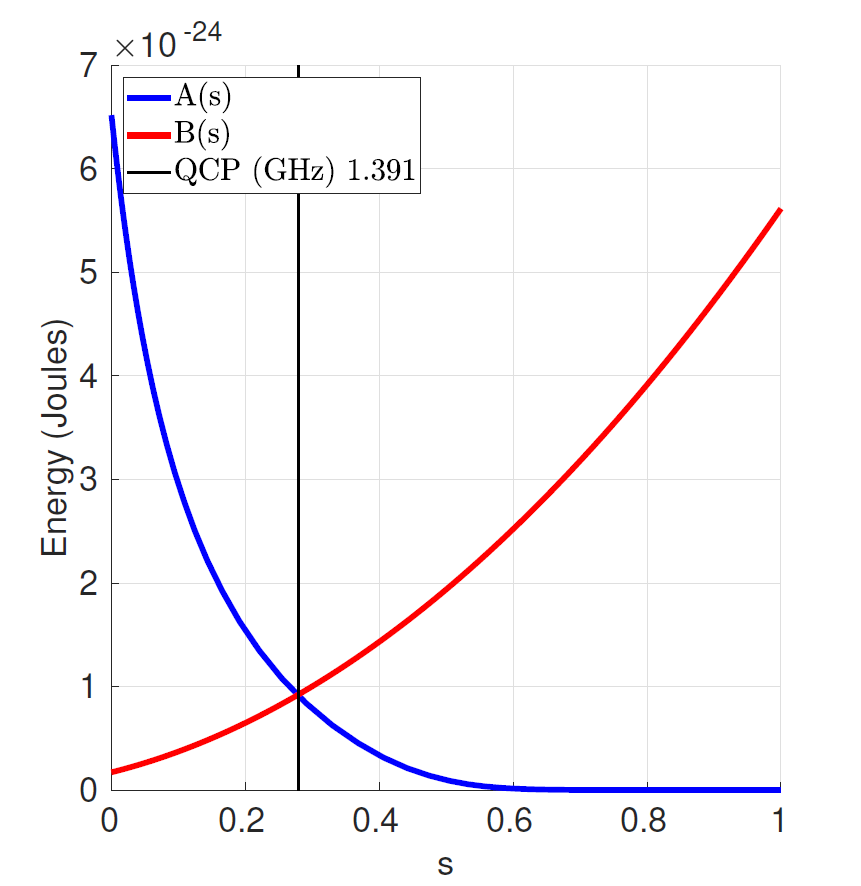}}
\hfill
\subfloat[The qubit topology (Pegasus)]{\label{fig: pegasus} \includegraphics[width=0.45\textwidth]{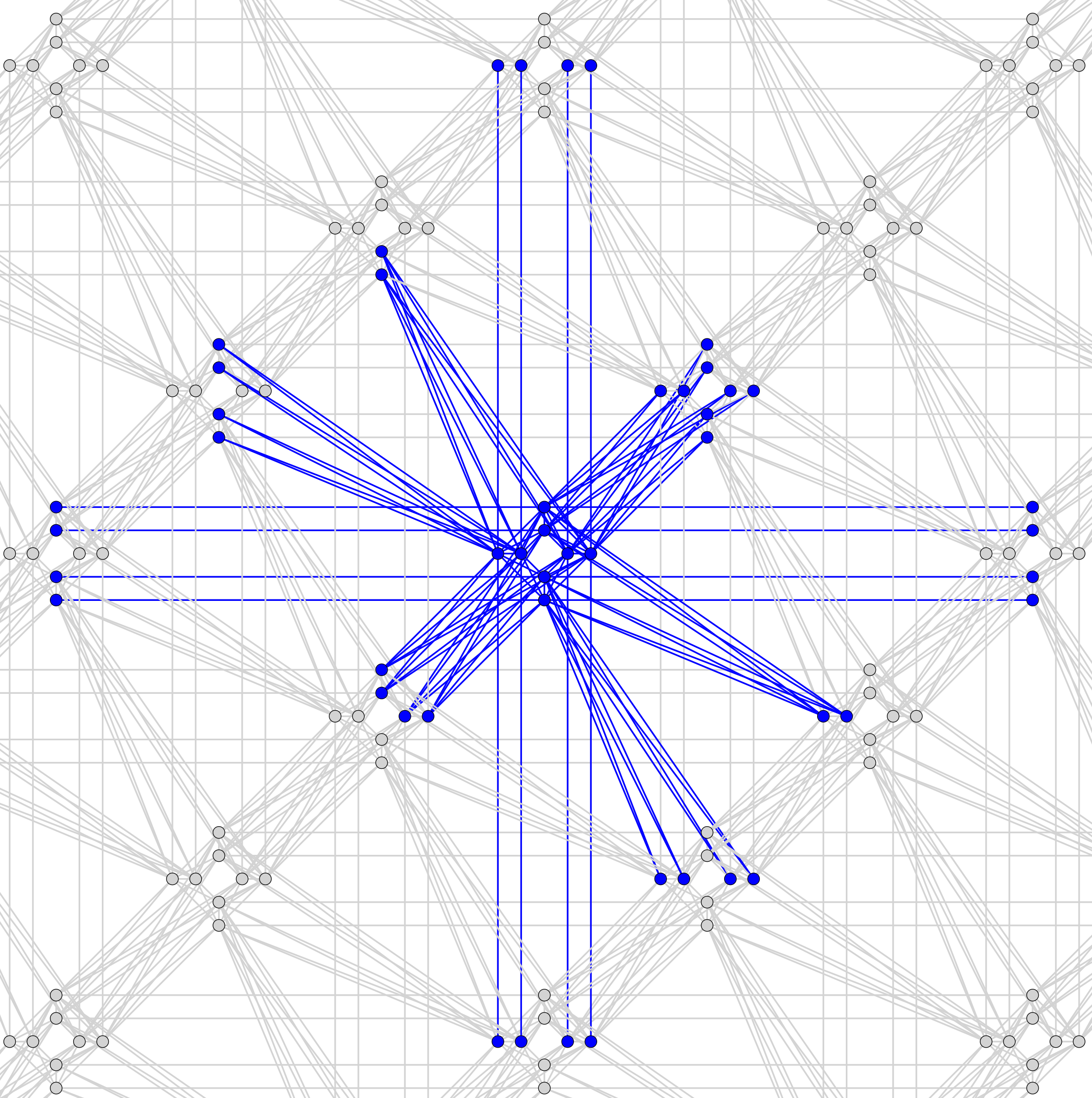}}%
\label{4figs-bis}
\caption{Information about the D-Wave Pegasus systems and their behavior.}
\end{figure}

\ignore{
\begin{figure}[!tbp]
\begin{minipage}{.5\linewidth}
\centering
\includegraphics[width=.6\textwidth]{figs/annealing-schedule.png}
\caption{\label{fig: annealing-schedule} Annealing Schedule}
\end{minipage}
\begin{minipage}{.5\linewidth}
\centering
\includegraphics[width=.6\textwidth]{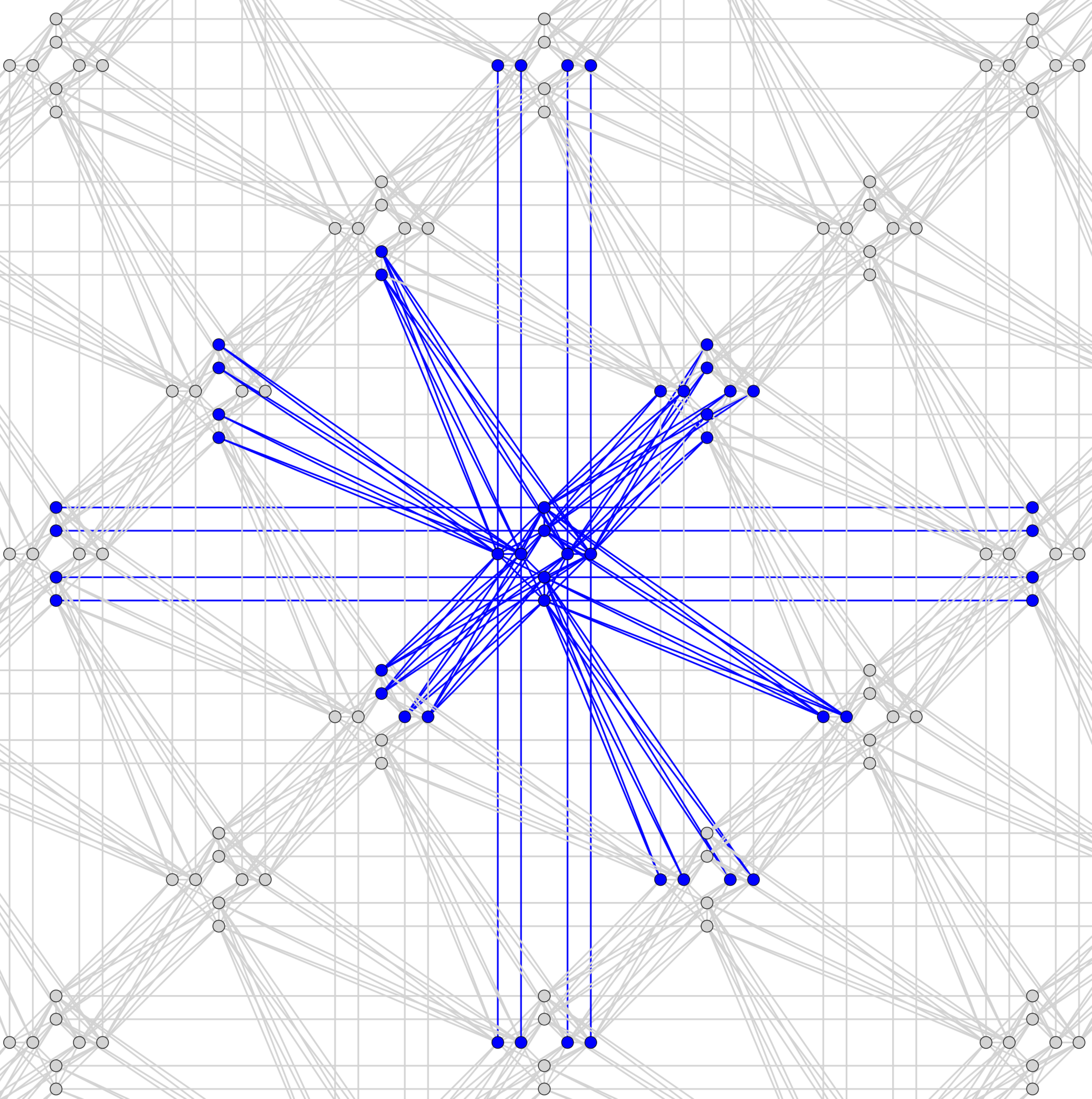}
\caption{\label{fig: pegasus} The qubit topology (Pegasus)}
\end{minipage}
\end{figure}
}

\subsection*{Monolithic encoding of small SAT problems based on OMT}

 Bian et al.\cite{Bian2020} formulated the problem of
 encoding 
 SAT problems into {\it Ising models}
 that are compatible with the available quantum topology ---represented
 as a graph  $(V, E)$ s.t. the nodes $V$ are the qubits and the edges
 $E$ are the qubit couplings--- with the goal of feeding them to the quantum annealer.
 Here we briefly summarize their techniques, adopting the same notation.

 Given a (small enough) Boolean formula $\Fx$ and a set of extra
 Boolean variables $\as$ (called {\em ancillae}), we first need
 to map
 the Boolean variables \xs and \as into a subset
 $\zs\subseteq V$ of 
 the qubits in the topology, with the intended meaning that the qubit
 values $\set{1,-1}$ are interpreted as the truth
 values $\set{\top,\bot}$ respectively.
 (With a little abuse of notation, we consider this map implicit
 and say that $\zs\defas\xs\cup\as$.) This map, called {\em placement},  can be performed either manually or via
 ad-hoc procedures\cite{Bian2020}.
 
 Then we need to compute the values $\theta_0$, $\theta_i$, and
 $\theta_{ij}$ of a {\it penalty function} $\Pxa$ such that, for some
 value $g_{min}>0$:
 \newcommand{\Pxaz}{\ensuremath{P_F(\underbrace{\xs,\as}_{\zs}|\ts)}\xspace}
\begin{eqnarray}\label{eq:penfunction-ancillas}\textstyle
\Pxaz \defas \offset{} + \sum_{\substack{z_i\in V}} \theta_{i} z_i +
\sum_{\substack{(z_i,z_j)\in E, i<j}}
  \theta_{ij} z_i z_j;
  \ z_i \in \{-1, 1\};
  & & \text{ } \ 
      \forall \xs\ \ min_{\set{\as}} \Pxa
  \begin{cases}
= 0 &\text{ if }   F(\xs)=\top \\
   \geq g_{min} &\text{ if } F(\xs)=\bot
  \end{cases}
\end{eqnarray}
 Intuitively, \Pxa allows for discriminating truth values for $\xs$ which
satisfy the original formula \Fx (i.e., these s.t. $min_{\set{\as}} \Pxa=0$) from
these who do not (i.e., these s.t. $min_{\set{\as}} \Pxa\ge g_{min}$).
$\theta_0$, $\theta_i$, $\theta_{ij}$ and $g_{min}$ are called respectively {\em
  offset}, {\em biases},  {\em couplings} and the {\em gap};
 the offset has no bounds, whereas biases and couplings have a fixed range of possible values
 ($[-2,+2]$ for biases and $[-1,+1]$ for coupling for the old Chimera
 architecture, $[-4,+4]$ for biases and $[-2,+1]$ for couplings for
 the Pegasus architecture of Advantage systems).

The penalty function \Pxa{} \eqref{eq:penfunction-ancillas} is fed to
the quantum annealer,  which tries to find values for the \zs's which
minimizes it. Once the annealer  reaches a final
configuration, if the corresponding energy is zero, then we can 
conclude that the original formula is satisfiable and the values of
$\xs\subseteq\zs$ 
satisfy \Fx ---once reconverted from  $\set{1,-1}$ to  $\set{\top,\bot}$.  
Notice that we may have a solution for $\Fx$ even if the energy of the assignment is not zero, because 
the truth values of the ancillae do not impact the satisfiability of
the original formula $\Fx$ but may affect the final energy. (We will call
them ``$>0$-energy solutions''.)
This is not an issue, because checking if the truth
assignments of the variables in $\xs$ satisfy $\Fx$ is
trivial. Notice also that, since the annealer is not guaranteed to
find a minimum, if the result is not a solution, then we cannot conclude
that \Fx is unsatisfiable. 

 The gap $g_{min}$ between  ground and non-ground states has a fundamental role in
making the annealing process more effective: the bigger $g_{min}$, the easier is
for the annealer to discriminate between satisfying and non-satisfying
assignments. 
Ancillae \as{} are needed to increase the number of $\theta$ parameters, because the problem of
finding a suitable \Pxa matching \eqref{eq:penfunction-ancillas} is
over-constrained in general, so that without ancillae there would be no
penalty function even for very few variables \xs's (e.g., $>3$). 
The more ancillae, the more degrees of freedom, the higher the chances
to have a suitable penalty with a higher gap $g_{min}$.

The problem of synthesizing $\Pxa$ is solved by using a solver for
{\em Optimization Modulo Theories} such as OptiMathSAT
\cite{Sebastiani2020}. For the Pegasus architecture, we feed
OptiMathSAT some formula equivalent to:
\begin{eqnarray}
\label{eq:encoding1}
&&
\ \forall \xs. 
    \left [ 
    \begin{array}{ll}
      {(\pos\Fx \imp \exists \as. (\Pxa=0 ))\  \wedge}\\
      (\pos\Fx \imp \forall \as. (\Pxa\ge 0 ))\  \wedge\\
      (\neg\Fx \imp \forall \as. (\Pxa\ge g_{min} )) \wedge\\
      \bigwedge_i(\theta_i \in [-4,4]) \wedge \bigwedge_{i,j}(\theta_{ij} \in [-2,1]) 
      
    \end{array}
    \right ],
  \end{eqnarray}
asking to find the set of values of the $\theta$s satisfying
\eqref{eq:encoding1} which maximizes the gap $g_{min}$. The result, if
any, is a suitable \Pxa.

\subsection*{Locally-structured embedding for large SAT problem}
Encoding a Boolean formula $F(\xs)$ using the monolithic encoding
shown in (\ref{eq:encoding1}) presents several limitations.
In practice, no more than
10 qubits can be considered if we directly use the formulation in
equation (\ref{eq:encoding1}), and recalling that some of them are
required as ancillary variables, the set of Boolean formulas we can
encode monolithically this way is quite limited.  

To encode larger propositional problems, Bian et al.\cite{Bian2020} proposed a {\it divide-and-conquer}
strategy. The original formula is first And-decomposed into smaller sub-formulae
so that the penalty function $\Pxa$ for each subformula can be
computed for some given placement. In particular,
given a formula $\Fx$, we can And-decompose it as
$F(\xs):= \bigwedge_{k=1}^K F_k(\xs^k)$, s.t. each penalty function can
be computed offline by {\sf OptiMathSAT}. The {\em And-decomposition
property}\cite{Bian2020} guarantees under some conditions that the penalty function of the original formula
$\Fx$ can be easily obtained by summing up all the penalty functions
from the subformulae: $\Pxa = \sum_k \Pxak$, where $g_{min}(\Fx) =
min_k(g_{min}^k(F_k(\xs)))$. 
\ignore{
  The decomposition task is performed through DAG-aware minimization by
using the logic synthesizer ABC.
}
The penalty function $P_{F_k}(\xsk,\ask|\ts^k)$ of each sub-formula $F_k(\xs^k)$ is then
mapped into a subgraph in the QA topology --e.g. one of the tiles in
the Pegasus topology. 

 When two sub-formulae $F_i(\xs^i)$ and
 $F_j(\xs^j)$ share one (or more) Boolean variables $x$,
 we can (implicitly) rename one of the two occurrences  into
 $x'$
 and   conjoin a chain of equivalences  $x\iff ... \iff x'$ to them.
 (I.e., $F_i(...,x,...)\wedge F_j(...,x,...)$ can be (implicitly) rewritten into
 $F_i(...,x,...)\wedge F_j(...,x',...)\wedge(x\iff ... \iff x')$.)
 This corresponds to linking the corresponding qubits $x$ and $x'$ in the penalty
 functions \Pxai and \Pxaj 
 by means of a {\it
   chain} of unused qubits used as ancillary
 variables, forcing all involved qubits to assume the
 same truth value, by using the {\it equivalence chain penalty
   function} $\sum_{(z, z') \in chain} (2 - 2zz')$ for the qubits in
 the chain,   corresponding to
 the Boolean formula $x \iff ...\iff x'$ (here we
 consider the Pegasus extended ranges).  The final penalty function
 is the sum of the 
 penalty functions from the decomposition phase with those of the chains.  

 We refer the reader to Bian et al.\cite{Bian2020} for a more detailed
 description of these techniques.

%% file: encoding.tex
\section*{Encoding binary multipliers into Pegasus quantum annealers}
\subsection*{Modular representation of a multiplier}

\begin{figure}
\centering
\subfloat[The theoretical idea behind a $4\times 4\text{-bit}$ shift-and-add multiplication]{\label{fig: arith_multiplier}{
\includegraphics[width=0.45\textwidth]{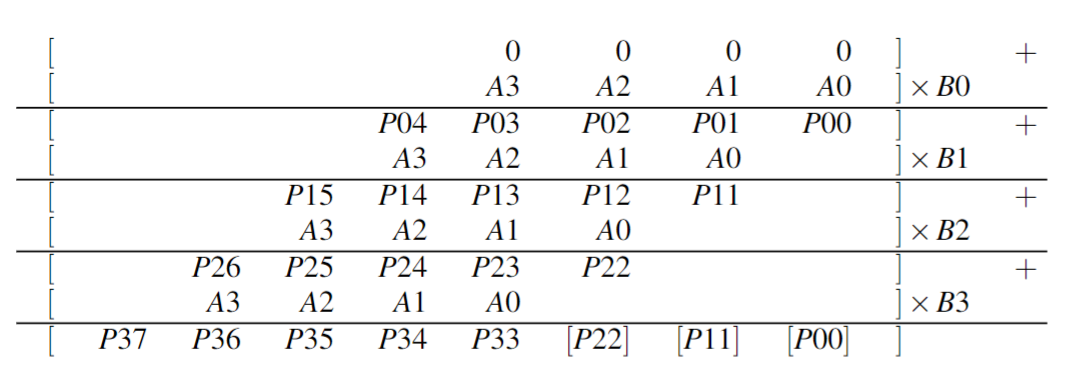}}
}
\hfill
\subfloat[The $4\times 4\text{-bit}$ multiplier schema of Figure \ref{fig: arith_multiplier}]{\label{fig: multiplier} \includegraphics[width=0.53\textwidth]{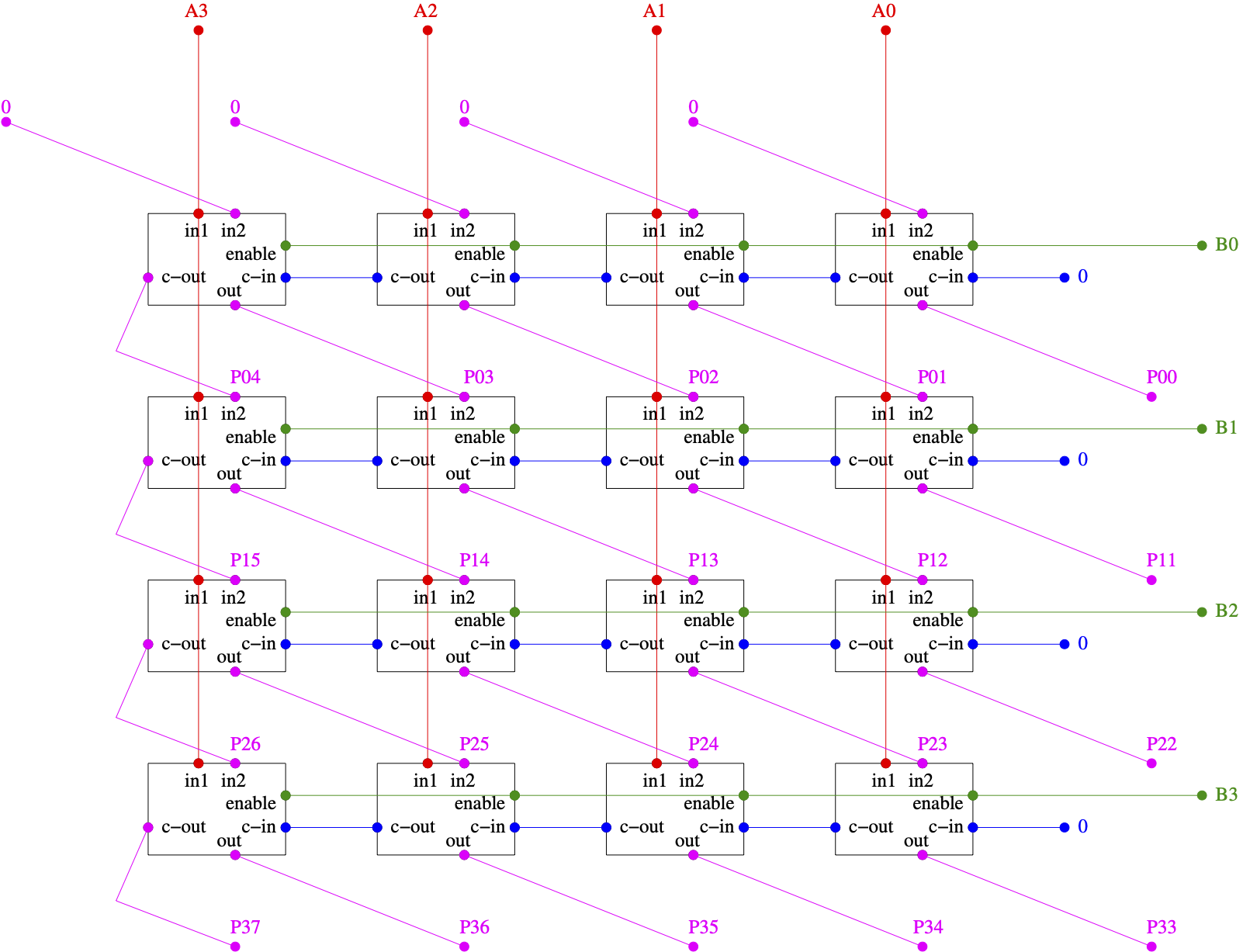}}%
\label{4figs-quat}
\caption{Details about the modularity of shift-and-add multipliers.}
\end{figure}

\ignore{
\begin{figure}[!tbp]
\footnotesize
\begin{minipage}{.41\textwidth}
\scriptsize{ 
$$
\begin{array}{rrrrrrrrrlrrrrrr}
\ [&  &  &  &  &  0 &  0 &  0 &  0 & ]& + \\
\ [&  &  &  &  & A3 & A2 & A1 & A0 & ]\times B0 &  \\
\hline
\ [&  &  &  &P04&P03&P02 &P01 &P00 & ]& + \\
\ [&  &  &  & A3 & A2 & A1 & A0 & & ]\times B1 & \\
\hline
\ [&  &  &P15&P14&P13 &P12 &P11 & & ]& + \\
\ [&  &  & A3 & A2 & A1 & A0 & & &  ]\times B2 & \\
\hline
\ [&  &P26&P25&P24 &P23 &P22 & & & ]&   + \\
\ [&  & A3 & A2 & A1 & A0 & & & &   ]\times B3 & \\
\hline
\ [&P37&P36&P35 &P34 &P33 & [P22] & [P11]& [P00]&]&      \\
\end{array}
$$}
\caption{
A $4\times 4\text{-bit}$ shift-and-add multiplication: \mbox{$ [P37\ P36\ P35\ P34\ P33\ P22\ P11\ P00]$} \newline
\mbox{$=[A3\ A2\ A1\ A0]\times\ [B3\ B2\ B1\ B0]$}.
}
\label{fig: arith_multiplier}
\end{minipage}%
\begin{minipage}{.6\textwidth}
\centering
  \includegraphics[width=.8\linewidth]{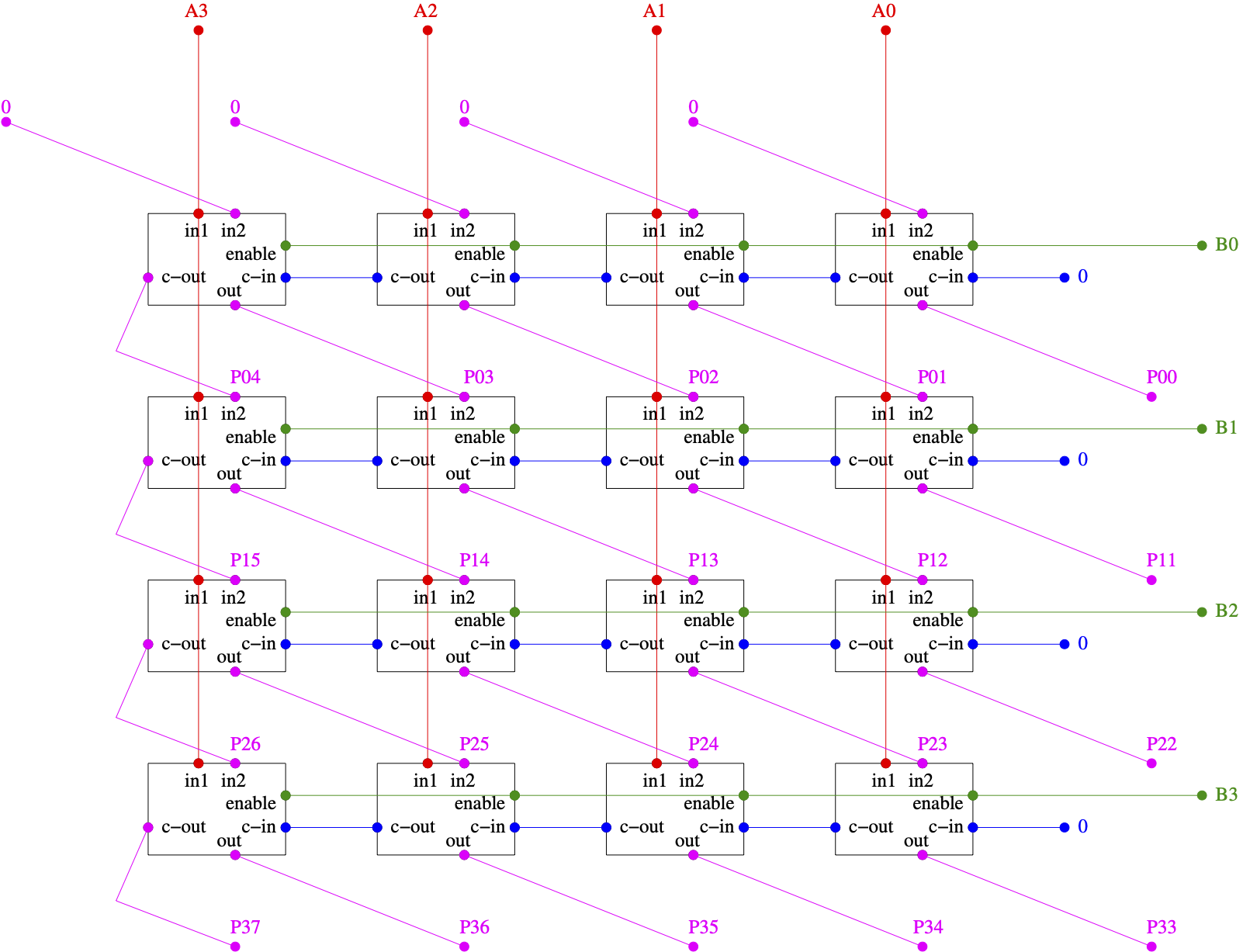}
  \caption{\label{fig: multiplier}
  The $4\times 4\text{-bit}$ multiplier schema of Figure \ref{fig: arith_multiplier}.}
\end{minipage}
\end{figure}
}

In a fashion similar to Bian et al. \cite{Bian2020}, we developed a
{\em modular} encoding of a shift-and-add multiplier, so that it could be
easily extended for future larger quantum devices. To this extent, the
binary-arithmetic computation of multiplications, as shown in Figure
\ref{fig: arith_multiplier}, is based on a module implementing a  {\em Controlled Full-adder
  (CFA)}.
The Boolean representation of a single CFA is:
\begin{align}
CFA (in2, in1, enable, c\_in, c\_out, out) \defas 
	     & \big(c\_out \leftrightarrow ((c\_in \land ((enable \land in1) \lor in2)) 
	\lor ((enable \land in1) \land in2)\big)  \nonumber\\
	\land& \big(out \leftrightarrow ((enable \land in1) \oplus in2 \oplus c\_in)\big) \nonumber
 \label{eq: cfa}
\end{align}
The structure of a CFA includes four inputs: two operand bits ($in1$
and $in2$), a control bit $(enable)$ and a carry-in bit $c\_in$.
The
output-carry bit $c\_out$ and the output $out$ of a CFA are computed
as is it typically done for classical full adder, the only difference
being the the fact that the input $in1$ is enabled by the $enable$
bit: 
when $enable$ is
true, the CFA behaves as a standard full adder; when $enable$
is false, the CFA behaves as if $in1$ were false.

As  shown in Figure \ref{fig: multiplier}, 
 an $m\times n$-bit multiplier can be encoded using $m\cdot n$ CFAs as follows: 
\begin{eqnarray}
    \label{eq:multip}
	F_{P=A\times B} &= \bigwedge_{i=0}^{n-1}\bigwedge_{j=0}^{m-1}  
	CFA (in2^{(i, j)}, in1^{(i, j)}, enable^{(i, j)}, c\_in^{(i, j)}, c\_out^{(i, j)}, out^{(i, j)}) \land \bigwedge_{(x, x') \in chains} (x \leftrightarrow x')
\end{eqnarray}
where $chains$ corresponds to the set of all the equivalence chains
corresponding to the links between bits belonging to different CFAs, as in Figure \ref{fig:
  multiplier} (e.g. $(enable^{(i,j)}\iff enable^{(i,j+1)}$).

\subsection*{LSE-based encoding with qubit sharing, virtual chains, and alternating CFAs}

\begin{figure}
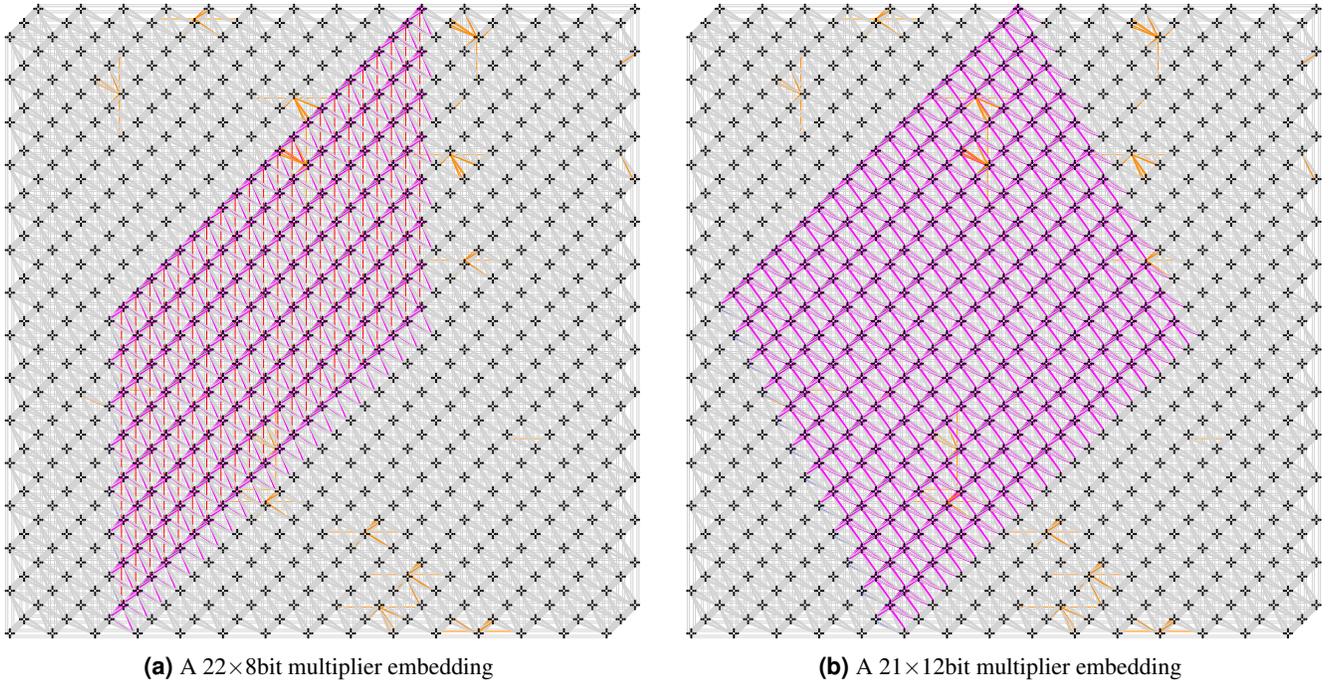

\centering
\subfloat[A 22$\times$8bit multiplier embedding]{\label{fig: multiplier_ver2}
\includegraphics[width=0.48\textwidth]{figs/22-8bit_multiplier_ver2_to_link_B_4_paper.pdf}}
\hfill
\subfloat[A 21$\times$12bit multiplier embedding]{\label{fig: multiplier_ver3} \includegraphics[width=0.48\textwidth]{figs/21-12bit_multiplier_ver3_to_link_B_4_paper.pdf}}%
\label{4figs-tris}
\caption{Modular encoding of binary multipliers on the D-Wave Pegasus topology.}
\end{figure}

\ignore{
\begin{figure}[!tbp]
	\begin{minipage}{.5\linewidth}
			\includegraphics[width=1\linewidth]{figs/22-8bit_multiplier_ver2_to_link_B_4_paper.pdf}
		\caption{A 22$\times$8bit multiplier embedding}
		\label{fig: multiplier_ver2}
	\end{minipage}
	\begin{minipage}{.5\linewidth}
			\includegraphics[width=1\linewidth]{figs/21-12bit_multiplier_ver3_to_link_B_4_paper.pdf}
		\caption{A 21$\times$12bit multiplier embedding}
		\label{fig: multiplier_ver3}
	\end{minipage}
\end{figure}
}


A direct approach to building multipliers using multiple CFAs is to
encode each CFA into a single Pegasus tile, using 2 of the 8 total
qubits as ancillae. Once the penalty function for a single CFA has been obtained, we can embed them modularly and generate a grid of CFAs that simulates the multiplier. Since some qubits are shared
among different CFAs, we must add equivalence chains to force
the equality of the values of the corresponding qubits.
First, the carry-out $c\_out$ qubit of a CFA placed into one tile
must be linked to the carry-in $c\_in$ qubit of the CFA placed in the
tile hosting the left CFA in the grid in
Figure~\ref{fig: multiplier}.
The same applies to the output $out$ of a CFA
and the input $in2$ in the bottom-left CFA in
Figure~\ref{fig: multiplier}.
Lastly, it is necessary
to generate the qubits links corresponding to the long red vertical chain and
the green horizontal chain in
Figure~\ref{fig: multiplier}, linking respectively the $in1$ and
$enable$ bits.

In the Pegasus topology, each tile has some direct connections with the neighbor
tiles along several directions (expressed in degrees counterclockwise
wrt. the horizontal line): $0^\circ$, $90^\circ$, $45^\circ$,
$120^\circ$ and $150^\circ$. 
Considering all these constraints, two macro-configurations for
placing the CFA grid of Figure~\ref{fig: multiplier} into a Pegasus architecture can be considered.
In both configurations, due to the high number of inter-tile $45^\circ$
connections, the horizontal connections in Figure~\ref{fig:
  multiplier} (the $c\_out-c\_in$ and $enable$ links) are placed
along the $45^\circ$ inter-tile connections.
With the first configuration, in Figure \ref{fig: multiplier_ver2}, the input
qubits $in1$ from vertically aligned CFAs in the grid are connected by
90$^\circ$ inter-tile connections and the $out-in2$ links are
connected via $120^\circ$ ones. This allows for
fitting a $22 \times 8$-bit
multiplier into the whole Pegasus topology.
The second configuration, in Figure \ref{fig: multiplier_ver3}, differs from the first one by chaining
the  $in1$ qubits along 120$^\circ$ connections and the
$out-in2$ links along 150$^\circ$ ones. Using diagonal
chains has the main advantage to fit a larger $21 \times 12$-bit multiplier.
Both
configurations work modulo symmetries: for instance, encoding the grid
of CFAs such that the input variable $in1$ is propagated bottom-up
instead of top-down is feasible by slightly changing the qubits
placement into the tile.

Unfortunately, an 8-qubit CFA encoding to replicate the
two configurations described above turned out to be unfeasible in practice, because
no such encodings can be generated. This fact is due to two main
issues:
($i$) the low number of ancillae (only 2) available for encoding each
CFA, which drastically reduces the chances of finding a suitable penalty
function, and
($ii$) the absence of pairwise
direct 45$^\circ$ couplings between the same qubits in the neighbor
tiles, which prevents any direct implementation of the $enable$ chain
along the 45$^\circ$ direction.
(A similar issue occurs also in the second macro-configuration of
Figure~\ref{fig: multiplier_ver3} for the the $in1$ bit along the
120$^\circ$ direction.)
%
\\To cope with these issues, we propose three novel techniques: {\it Alternating CFAs}, {\it Qubit sharing}, and {\it Virtual chaining}. 

\paragraph{Alternating CFAs. }

To address the issue (ii) of missing couplings between qubits on the
45$^\circ$ direction, we propose to alternate two slightly-different
CFAs in tiles along the 45$^\circ$ line. 
In particular, in Figure \ref{fig: CFA0_encoding} and \ref{fig: CFA1_encoding} we make the OMT solver compute two different CFAS
forcing $enable$ to be positioned respectively in the first vertical qubit on the upper tile and the
third horizontal qubit in the 45$^\circ$-degree bottom-left
tile. Such qubits are pairwise directly coupled, allowing thus a chain
for $enable$ qubit along the 45$^\circ$-degree direction (the green
links). 
We stress the fact that the two different CFA encodings are not guaranteed to
have the same gap $g_{min}$, and that different placements leading to
different $g_{min}$ values typically may negatively affect the annealing
process.   

\paragraph{Qubit sharing.}

\begin{figure}
\centering
\subfloat[V1.]{\label{fig: CFA_encoding_v1}
\includegraphics[width=0.22\textwidth]{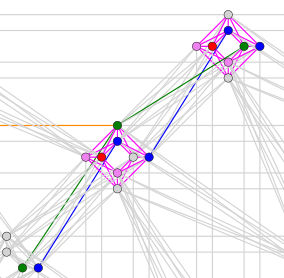}}
\hfill
\subfloat[V2.]{\label{fig: CFA0_encoding} \includegraphics[width=0.22\textwidth]{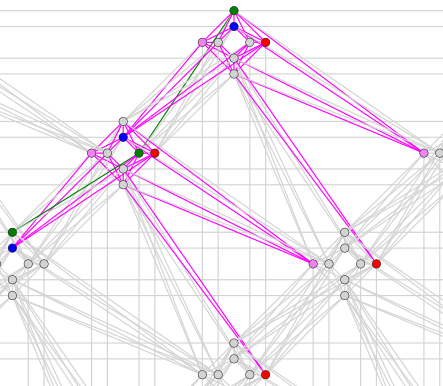}}%
\hfill
\subfloat[V3.]{\label{fig: CFA1_encoding}
\includegraphics[width=0.22\textwidth]{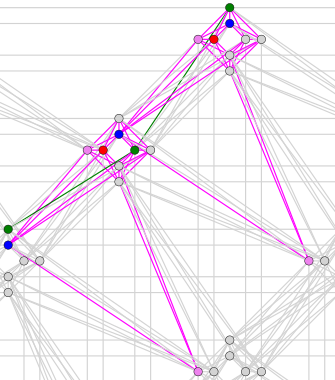}}%
\hfill
\subfloat[V4.]{\label{fig: unify}
\includegraphics[width=0.22\textwidth]{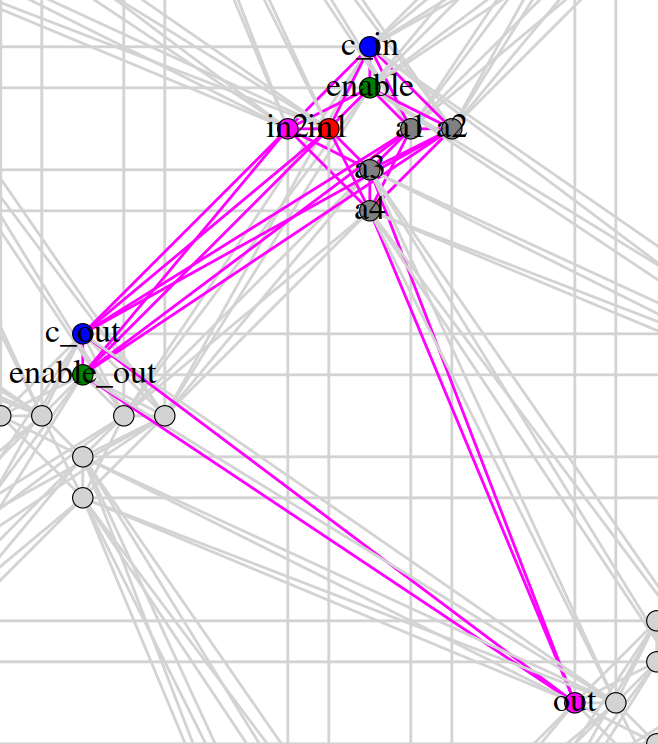}}%
\caption{CFA structure for the four versions of multipliers.}
\label{4figs}
\end{figure}

\ignore{
\begin{figure}[!tbp]
\begin{minipage}{.24\linewidth}
	\centering
	\includegraphics[width=.9\linewidth]{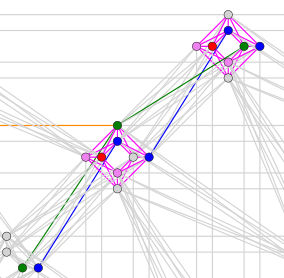}
	\caption{\label{fig: CFA_encoding_v1}  Alternating CFAs topology for the 22$\times$8 multiplier}
\end{minipage}
\begin{minipage}{.25\linewidth}
	\centering
	\includegraphics[width=.9\linewidth]{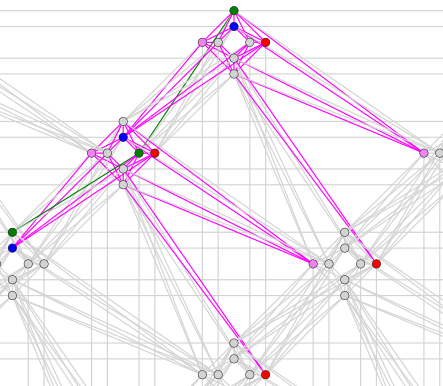}
	\caption{\label{fig: CFA0_encoding}  Alternating CFAs topology for the 21$\times$12 multiplier}
\end{minipage}
\begin{minipage}{.24\linewidth}
	\centering
	\includegraphics[width=.9\linewidth]{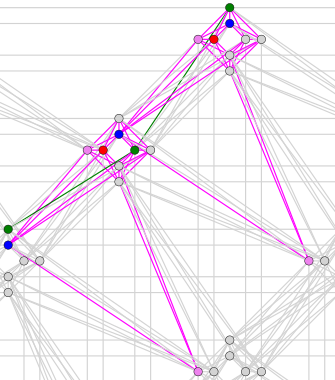}
	\caption{\label{fig: CFA1_encoding}  Alternating CFAs topology for the 22$\times$8 multiplier}
\end{minipage}
\begin{minipage}{.25\linewidth}
	\centering
	\includegraphics[width=.9\linewidth]{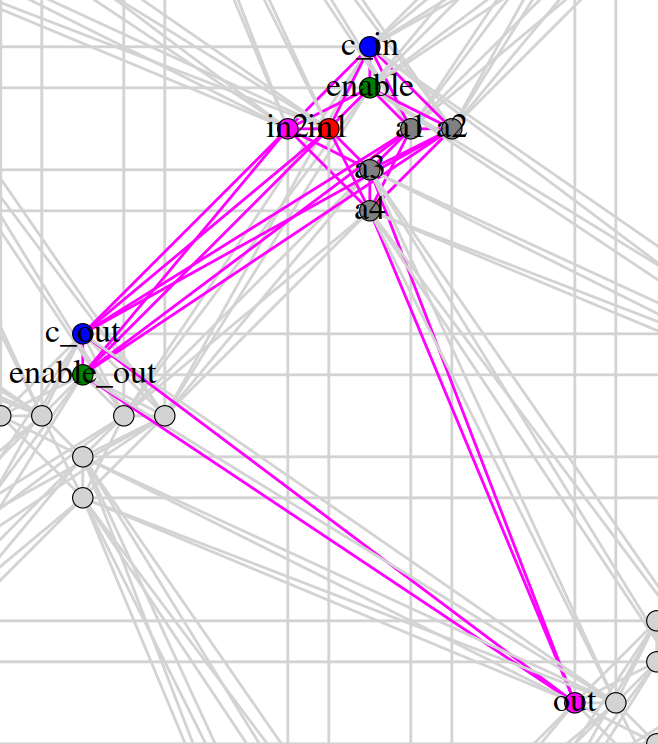}
	\caption{\label{fig: unify} Unified CFA embedding.}
\end{minipage}
\end{figure}
}

\ignore{
\begin{figure}[!tbp]
\begin{minipage}{.24\linewidth}
	\centering
	\includegraphics[width=.9\linewidth]{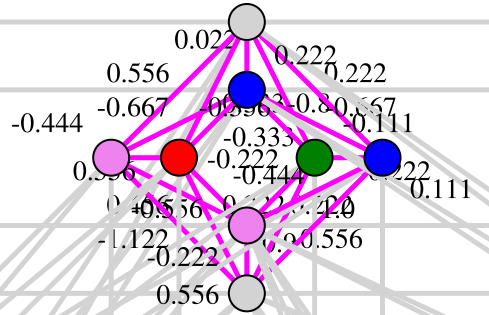}
	\includegraphics[width=.9\linewidth]{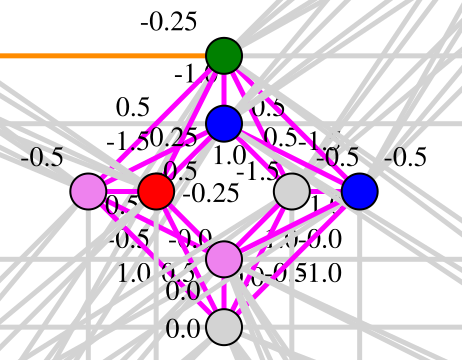}
\end{minipage}
\begin{minipage}{.24\linewidth}
	\centering
	\includegraphics[width=.9\linewidth]{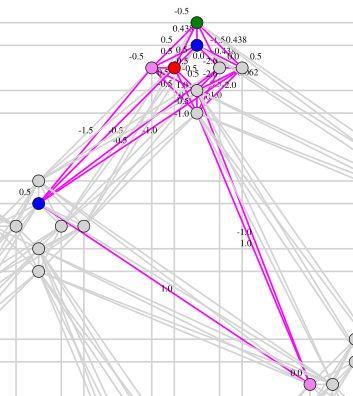}
	\includegraphics[width=.9\linewidth]{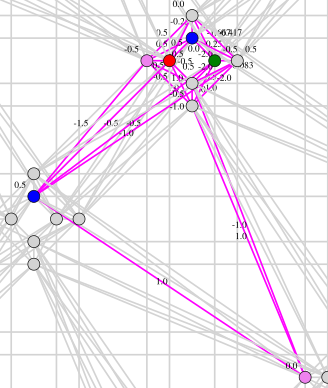}
\end{minipage}
\begin{minipage}{.25\linewidth}
	\centering
	\includegraphics[width=.9\linewidth]{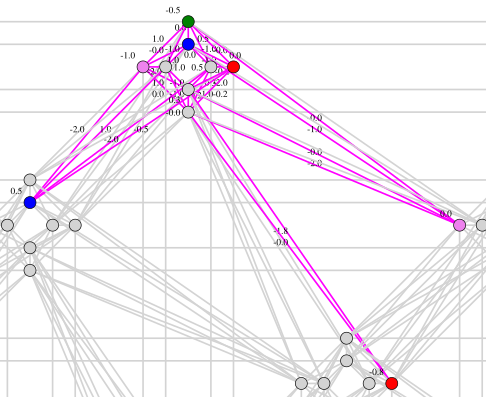}
	\includegraphics[width=.9\linewidth]{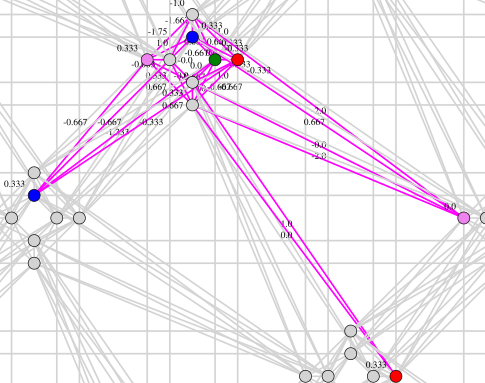}
\end{minipage}
\begin{minipage}{.25\linewidth}
	\centering
	\includegraphics[width=.9\linewidth]{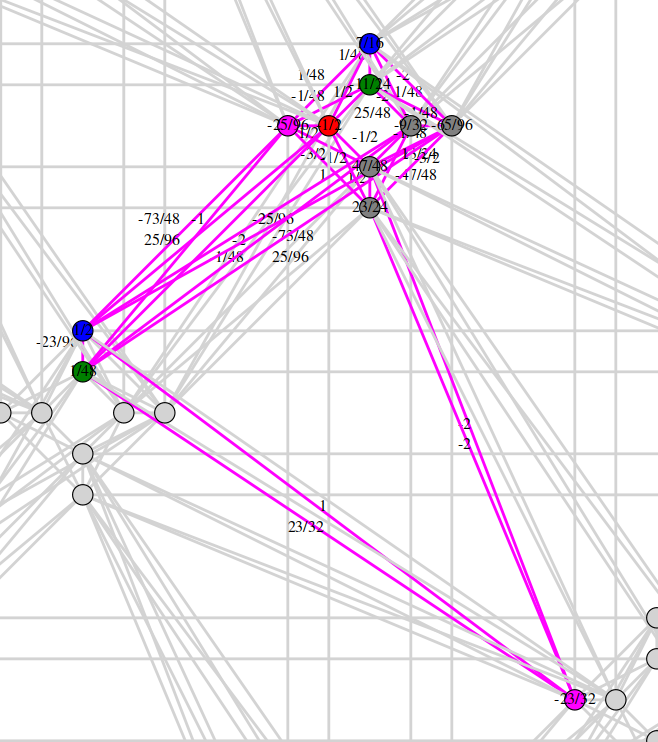}
\end{minipage}
\end{figure}
}

To address the issue (i) of the low number of ancillae, we propose a
technique to {\em share qubits between neighboring tiles}.
Rather than connecting two qubits from different CFAs with an
equivalence chain, we suggest utilizing a single qubit that is {\em
  shared} between the two CFAs.  
This means that the qubit will be used for the encoding of one CFA as
an output variable and as an input variable for the subsequent
CFA. This approach leads to partially-overlapping CFAs and the extra
qubit can be used as an ancillary variable to increase the minimum gap
of each CFA. 
Consider the schema in Figure~\ref{fig: unify}. The encoding of each CFA involves not only the 8 qubits of its tile but also the 3 qubits of neighbor tiles. 
In particular, the carry-out $c\_out$ is placed on the same qubit as
the carry-in $c\_in$ of the next 45$^\circ$-degree bottom-left tile
--corresponding to the left CFA in Figure~\ref{fig: multiplier}-- and
the $out$ qubit is placed in the same qubit of the $in2$ of the next
bottom-right 120$^\circ$-degree tile --corresponding to the lower-right
CFA in Figure~\ref{fig: multiplier}. The same idea applies also to the
schemata in Figures \ref{fig: CFA0_encoding} and \ref{fig: CFA1_encoding}.
(The role of the $enable\_out$ qubit in Figure~\ref{fig: unify} will be explained later.)

Notice that, since the global penalty function is the sum of the
penalty functions of all CFAs plus these of all the equivalence
chains, the value of the bias for the shared qubit in the global
penalty function is the sum of these two qubits with different
roles in the two penalty functions of the two sharing CFAs. (E.g., the
bias of the qubit which is a $c\_out$ for one CFA and a $c\_in$ for
another CFA is the sum of the $c\_in$ and $c\_out$ biases of a CFA
encodings.) 
Thus, to generate penalty functions for the CFAs that allow qubit
sharing, we introduce additional constraints to the OMT formulation
 in \eqref{eq:encoding1}. In particular, we add an arithmetical
constraint to force the sum of the biases of the shared qubits from
two CFAs to fit in the bias range, thus simulating their
over-imposition (e.g., we add a constraint like $(\theta_{c\_{in}} +
\theta_{c_{out}} \in [-4, 4])$). In fact, if the final bias values did
not fit into the range, then the D-Wave encoders would automatically
rescale all values of biases and couplings, reducing the $g_{min}$
value and thus negatively affecting the probability of reaching a global
minimum.  

\paragraph{Virtual chaining.}

The concept of qubit sharing can be exploited to simulate the
existence of equivalence links when physical connections are missing,
providing another solution to issue (ii). Consider the
CFA encoding in Figure \ref{fig: unify} and the $enable$ logical
variable. Its truth value is shared by all CFAs belonging to the same
row in the grid so that all the $enable$ qubit of each CFA should be
connected by an equivalence chain with the $enable$ qubit of the
45$^\circ$ bottom-left CFA. Unfortunately, there is no arc linking
pairwise the respective qubits of the tiles along this direction.

In such cases, two qubits that are intended to hold the same truth value but lack a direct coupling can be {\em virtually chained} by using the links with the common neighbors.
This is performed by extending the encoding as follows: 
\\(a) create a new virtual logical variable (i.e. $enable\_out$) to be placed in the qubit in the neighbor tile corresponding to the variable we want to chain virtually (i.e. $enable$); 
\\(b) extend the formula  defining a CFA by conjoining the equivalence constraint between the chained and the virtual variables (i.e., $CFA' (in2, in1, enable, c\_in, c\_out, out,enable\_out) \defas CFA (in2, in1, enable, c\_in, c\_out, out) \wedge (enable \iff enable\_out)$; 
\\(c) build the penalty function of $CFA'$ instead of $CFA$ by applying qubit-sharing also to $enable$ and $enable\_out$.

It should be noted that if two directly-connected qubits are both
involved in qubit sharing (i.e. $c\_in$ and $enable$), then also the
respective coupling is shared by the two CFAs. Therefore an arithmetic
constraint must be added to force the sum of the two couplings to be
in the coupling range
(i.e. $(\theta_{c\_in,enable}+\theta_{c\_out,enable\_out}\in
[-2,1])$).

\paragraph{Comparing different multiplier configurations.}
 Overall, exploiting Alternating CFAs, qubit sharing, and Virtual
chaining made it possible for us to generate four multiplier
configurations, which are summarized in Table~\ref{tb:three-confs}.
Versions V1, V3 and V4 allow for implementing the $22\times8$-bit
schema of Figure~\ref{fig: multiplier_ver2}, whereas version V2 allows for implementing the $21\times12$-bit
schema of Figure~\ref{fig: multiplier_ver3}.
Versions V2, V3 and V4 correspond to the encodings in
Figures~\ref{fig: CFA0_encoding}, \ref{fig: CFA1_encoding}
and~\ref{fig: unify} respectively.

In particular:  by exploiting {\em Alternating CFAs}, with versions V1, V2 and V3 (Figures ~\ref{fig: CFA_encoding_v1}, \ref{fig: CFA0_encoding}, \ref{fig: CFA1_encoding}),
we could implement an $enable$ chain along the 45$^\circ$ diagonal,
and with version V1 (Figure ~\ref{fig: CFA0_encoding}) an $in1$ chain along the 120$^\circ$ diagonal;
by exploiting
{\em Qubit sharing},  with versions V2, V3, V4 (Figures~\ref{fig: CFA0_encoding}, \ref{fig: CFA1_encoding} and~\ref{fig: unify}), we have 
saved two
qubits, which we could use as ancillae, improving also the quality of the
encodings and their gap $g_{min}$;
by exploiting {\em Virtual chaining},
with V4 (Figure~\ref{fig: unify}), we could implement a
virtual chain for the  $enable$ qubit along the 45$^\circ$ diagonal;
with V2 (Figure~\ref{fig:  CFA0_encoding}) we could implement a
virtual chain for the  $in1$ qubit along the 120$^\circ$ diagonal.

Version V1 (Figure~\ref{fig: CFA_encoding_v1}) implements the $22 \times 8$-bit macro-configuration of
Figure~\ref{fig: multiplier_ver2} and relies exclusively on
alternating CFAs, linking inter-tile qubits only by physical
chains. Although alternation allowed the production of an actual encoding,
which was not possible otherwise, without qubit sharing only two
ancillae were available, producing two alternating
configurations with different and very low gaps: $1$ and
$\frac{4}{9}$. These numbers are way lower than the gap used for
chains, the annealers tend to be stuck on local minima since
changing the spin of chained qubits becomes difficult.

Version V2 (Figure~\ref{fig: CFA0_encoding}) implements the
$21 \times 12$-bit macro-configuration of Figure~\ref{fig:
  multiplier_ver3} with alternating CFA encodings, using a virtual
chain for implementing the $in1$ chain along the 120$^\circ$
direction, and qubit sharing for the $c\_in-c\_out$ (the blue qubits)
and $out-in2$ (the magenta qubits) connections, which saves two qubits
and allows for 4 ancillae. This allows us to improve significantly the
gaps to $2$ and $\frac{4}{3}$ respectively. Nevertheless, the two CFAs
have different $g_{min}$, which negatively affects the global gap
(which is thus $\frac{4}{3}$) and thus the overall performances of the
annealer.

Version V3 (Figure~\ref{fig: CFA1_encoding}) instead implements the
$22 \times 8$-bit macro-configuration of Figure~\ref{fig:
  multiplier_ver2} with alternating CFA encodings, using a physical
90$^\circ$ $in1$, also using qubit sharing for the $c\_in-c\_out$ and
$out-in2$ connections, allowing 4 ancillae. With this configuration, we
obtain two CFAs with identical gap $2$, which is a significant
improvement.  Nevertheless, having two physical chains for two
different variables ($enable$ and $in1$) affects the annealer's
performances: the longer the chains, the more difficult is for the
quantum system to flip all values of the chained qubits and escape a
minimum.

Version
V4 (Figure~\ref{fig: unify}) also implements the $22 \times 8$-bit
macro-configuration of Figure~\ref{fig: multiplier_ver2}, but uses
only one CFA encoding of gap $2$.
This is achieved by exploiting not only qubit sharing for the $c\_in-c\_out$ and
$out-in2$ connections, but also virtual chaining for implementing the
$enable$ chain, whereas $in1$ is physically chained vertically.
By using a single CFA and having only one physical chain rather than
two, most of the issues affecting 
annealing in the previous cases is solved, thus the optimization of
the penalty function by the QA turns out to be more effective.
Consequently, all experiments in the subsequent section employ version V4. 
\begin{center}
\begin{table}[t]
\small
\centering
\begin{tabular}{c|cccc}
\hline
\textbf{ Multiplier version}                                            &\textbf{V1}
                                             &\textbf{V2}                      &\textbf{V3}                      &\textbf{V4}                                \\ \hline
Multiplier Max. Size & 22$\times$8                                   & 21$\times$12                               & 22$\times$8                               & 22$\times$8                                          \\ \hline
\# of ancillae per CFA & 2               & 4                                   & 4                                   & 4                                             \\ \hline
\# of different CFA encodings & 2               & 2                                   & 2                                   & 1                                             \\ \hline
Gap of CFA penalty functions & (1, $\frac{4}{9}$)                & (2,$\frac{4}{3}$)                               &  (2,2)                  & 2                                             \\ \hline
Connection $in1(i,j)-in1(i+1,j-1)$  & Chain (90$^\circ$)       & Virtual chain (120$^\circ$)  & Chain (90$^\circ$) & Chain (90$^\circ$)           \\ \hline
Connection $enable(i,j)-enable(i,+1)$  & Chain (45$^\circ$)      & Chain (45$^\circ$) & Chain (45$^\circ$) & Virtual chain (45$^\circ$) \\ \hline
Connection $c\_in(i,j)-c\_out(i,j+1)$  & Chain (45$^\circ$)       & Qubit sharing                             & Qubit sharing                             & Qubit sharing                                       \\ \hline
Connection $out(i,j)-in2(i+1,j-1)$ & Chain (45$^\circ$)        & Qubit sharing                             & Qubit sharing                             & Qubit sharing                                       \\ \hline
\end{tabular}
\caption{Comparison of the four multipliers obtained through qubit sharing and virtual chaining.}
	\label{tb:three-confs}
\end{table}
\end{center}

%% file: updated_results.tex
\ignore{
\subsection*{Initializing qubits -- continue and the improvement of CFA encoding}

To cope with the latter problem, we propose alternative methods 
to the $fix\_variables()$ function for initializing input qubits
in the encoded BQM.
They are ad-hoc encoding and extra-chain.
In the ad-hoc encoding, all the interface CFAs
that contains the input qubits required to initialize
are re-encoded by reduced embedding graphs offline,
with the coefficients of shared elements (qubits or couplings)
guaranteed to be in the restricted range for all combinations of inputs;
in the extra-chain, each input qubit are elicited an extra chain
and the initialization of the input qubit moves to the other endpoint.
This initialzation will not impact the encoding of interface CFA encoding,
due to the large range of qubit bias.
They are compared on solving small-size problems,
which are shown in the 'CFA0' column of Table \ref{tab: foward_annealing}.
Among the three initialization approaches, 
the ah-hoc encoding, which produces the new encodings 
for the interface CFAs with larger gaps $g_{min}$, 
outperforms the others, in terms of higher success probability to find groundstates,
whereas the extra-chain did not bring large improvement.

In order to further increase the success probability of finding grounstates,
we propose another initialization approach, flux-bias,
and also improve the encoding of CFA.
In this flux-bias, the input qubit is initialized through the flux bias of the qubit
from the hardware level, without touching the encoded BQM.
Specially, for a value $s_i\in\{-1,1\}$ to impose on a qubit, 
we set the flux bias for this qubit as 
$\phi_i = 1000\phi_0s_i$, where $\phi_0$ is
the default annealing flux-bias units of the DWave system 4.1
and the $1000$ is an empirical value we choose based on our experience.
On the other hand, the CFA encoding is further improved by 
minimizing the number of the first-excited states of BQM
on top of the found maximal $g_{min}=2$;
This improvement of CFA produces a BQM ('CFA1') that can function as CFA
for $98\%$ times, compared to the $90\%$ of the 'CFA0'.
Its improvement brought to factor integers is demonstrated 
by their comparsion on the extra-chain initialization.
This improved CFA encoding together with the flux-bias initialization
finally produce the highest success probability on factoring integers of 3+3, 4+4, 5+5 bits.
This improvment further enables to factor integers of up to 7+7 bits
with forward annealing ($T_a=10\mu s$).
Therefore, we continue to use this improved 'CFA1' BQM and flux-bias initialization
in the following sections to factor larger integers.
}

\ignore{
\begin{table}[ht]
\scriptsize
\begin{minipage}{.4\textwidth}
        \centering
        \begin{tabular}{|l|r|rrr|rr|}
            \toprule
           \multirow{2}{*}{size}     & \multirow{2}{*}{inputs}  & \multicolumn{3}{c}{CFA0} & \multicolumn{2}{c}{CFA1(improved)}\\
           & & api & ad-hoc & chain & chain & flux-bias\\
           \midrule
            \multirow{3}{*}{3*3}
            & 25  &	161 & 154 & 93      & 173 & 136   \\
            & 35  &	389 & 666 &	286     & 379 & 951   \\
            & 49  &	450 & 577 &	312     & 295 & 997   \\
            \midrule
            \multirow{3}{*}{4*4}
            & 121   & 17  & 4   & 30    & 33 & 0     \\
            & 143   & 40  & 52  & 28	& 32 & 67    \\
            & 169   & 31  & 54  & 4	    & 69 & 5     \\
            \midrule
            \multirow{15}{*}{5*5}
            & 289   & 5   & 0   & 0    & 1   & 0     \\
            & 323   & 2   & 0   & 1    & 3   & 0     \\
            & 361   & 1   & 1   & 0    & 1   & 3     \\
            & 391   & 6   & 1   & 4    & 19   & 9     \\
            & 437   & 17  & 0   & 3    & 2   & 0     \\
            & 493   & 3   & 6   & 0    & 0   & 2     \\
            & 527   & 21  & 11  & 6    & 5   & 37    \\
            & 529   & 5   & 0   & 3    & 1   & 8     \\
            & 551   & 0   & 11  & 4    & 3   & 4     \\
            & 589   & 16  & 13  & 11   & 22  & 52    \\
            & 667   & 0   & 6   & 2    & 9   & 105   \\
            & 713   & 11  & 12  & 3    & 1   & 138   \\
            & 841   & 5   & 9   & 8    & 8   & 7     \\
            & 899   & 17  & 76  & 5    & 13  & 343   \\
            & 961   & 1   & 43  & 0    & 0   & 338   \\
            \midrule
            \multirow{10}{*}{6*6}
            & 2257	 &    &    &        &   & 1         \\
            & 2501	 &    &    &        &   & 2         \\
            & 2623	 &    &    &        &   & 1         \\
            & 2773	 &    &    &        &   & 39        \\
            & 2867	 &    &    &        &   & 14        \\
            & 3127	 &    &    &        &   & 1         \\
            & 3233	 &    &    &        &   & 1         \\
            & 3481	 &    &    &        &   & 3         \\
            & 3599	 &    &    &        &   & 4         \\
            & 3721	 &    &    &        &   & 0         \\
            \midrule
            \multirow{10}{*}{7*7}
            & 10033	 &    &    &        &   & 0      \\
            & 10541	 &    &    &        &   & 1      \\
            & 11303	 &    &    &        &   & 0      \\
            & 12319	 &    &    &        &   & 0      \\
            & 12827	 &    &    &        &   & 1      \\
            & 13081	 &    &    &        &   & 2      \\
            & 13589	 &    &    &        &   & 10     \\
            & 13843	 &    &    &        &   & 0      \\
            & 14351	 &    &    &        &   & 0      \\
            & 16129	 &    &    &        &   & 7      \\
            \midrule
            \multirow{10}{*}{8*8}
            & 49447	 &    &    &        &   & 0   \\
            & 49949	 &    &    &        &   & 0   \\
            & 52961	 &    &    &        &   & 0   \\
            & 55973	 &    &    &        &   & 0   \\
            & 56977	 &    &    &        &   & 0   \\
            & 57479	 &    &    &        &   & 0   \\
            & 58483	 &    &    &        &   & 0   \\
            & 59989	 &    &    &        &   & 2   \\
            & 60491	 &    &    &        &   & 0   \\
            & 63001	 &    &    &        &   & 0   \\
            \bottomrule
        \end{tabular}
        \caption{\label{tab: foward_annealing} Forward annealing with $T_a=10\mu s$.}
\end{minipage}
\begin{minipage}{.6\textwidth}
        \centering
        \begin{tabular}[pos]{|l|r|r|r|r|}
            \toprule
            size & inputs & $S_p|T_p=100\mu s|T_a=10\mu s$ & $P_F$ & \#$(P_F=0)$ \\
            \midrule
            \multirow{10}{*}{8*8}
            & 49447     & 0.38	    & 0.0	    & 1     \\
            & 49949     & 0.51	    & 4.083	    & 0     \\
            & 52961     & 0.51	    & 6.0	    & 0     \\
            & 55973     & 0.33	    & 0.0	    & 6     \\
            & 56977     & 0.33	    & 0.0	    & 1     \\
            & 57479     & 0.33	    & 0.0	    & 3     \\
            & 58483     & 0.51	    & 6.083	    & 0     \\
            & 59989     & 0.33	    & 0.0	    & 43    \\
            & 60491     & 0.38	    & 0.0	    & 1     \\
            & 63001     & 0.51	    & 2.0	    & 0     \\
            \midrule
            \multirow{10}{*}{9*8}
            & 100273	& 0.51	    & 4.083	    & 0     \\
            & 101291	& 0.51	    & 8.083	    & 0     \\
            & 107399	& 0.51	    & 4.0	    & 0     \\
            & 113507	& 0.51	    & 8.083	    & 0     \\
            & 115543	& 0.33	    & 0.0	    & 1     \\
            & 116561	& 0.51	    & 6.0	    & 0     \\
            & 118597	& 0.51	    & 4.0	    & 0     \\
            & 121651	& 0.33	    & 0.0	    & 1     \\
            & 122669	& 0.51	    & 8.167	    & 0     \\
            & 127759	& 0.36	    & 0.0	    & 1     \\
            \midrule 
            \multirow{10}{*}{10*8}
            & 201137	& 0.51	    & 6.167	    & 0     \\
            & 203179	& 0.51	    & 8.0	    & 0     \\
            & 215431	& 0.51	    & 6.083	    & 0     \\
            & 227683	& 0.34	    & 0.0	    & 1     \\
            & 231767	& 0.51	    & 8.083	    & 0     \\
            & 233809	& 0.51	    & 8.0	    & 0     \\
            & 237893	& 0.51	    & 6.0	    & 0     \\
            & 244019	& 0.51	    & 6.25	    & 0     \\
            & 246061	& 0.51	    & 6.167	    & 0     \\
            & 256271	& 0.35	    & 0.0	    & 2     \\
            \bottomrule
        \end{tabular}
        \caption{\label{tab: thermal_fluctuations_in_FW} Thermal fluctuations in the forward annealing
        for factoring ten integers of 8+8, 9+8 and 10+8 bits.}

        \begin{tabular}{|l|r|rr|rrrr|}
        \toprule
        \multirow{2}{*}{size}   & \multirow{2}{*}{inputs}  & \multicolumn{2}{c}{Forward($Tp=100\mu s|T_a=10\mu s$)} & \multicolumn{4}{c}{Reverse annealing($Tp=100\mu s|T_a=10\mu s$)}\\
                                &                          & $S_p$ & $min(P_F)$                      & $S_p$ & $P_F$ & moves & \#$(P_F=0)$ \\
        \midrule
        \multirow{4}{*}{8*8}
        & 49949     & 0.5	 & 2.0      & 0.33	& 0.0	& 233	& 7     \\
        & 52961     & 0.35	 & 2.0      & 0.41	& 0.0	& 177	& 1     \\
        & 58483     & 0.33	 & 2.083    & 0.33	& 4.0	& 144   & 0      \\
        & 63001     & 0.51	 & 2.0      & 0.35	& 0.0	& 168	& 4     \\
        \midrule
        \multirow{5}{*}{9*8}
        & 107399    & 0.51	 & 4.0	    & 0.33	& 4.0	& 79    & 0      \\
        & 113507    & 0.38	 & 2.0	    & 0.33	& 2.0	& 71    & 0      \\
        & 116561    & 0.36	 & 4.0	    & 0.37	& 0.0	& 98	& 35    \\
        & 118597    & 0.33	 & 2.0	    & 0.33	& 4.0	& 201   & 0      \\
        & 122669    & 0.48	 & 4.083	& 0.36	& 0.0	& 129	& 7     \\
        \midrule 
        \multirow{5}{*}{10*8}
        & 231767   	& 0.33	 & 2.083	& 0.33	& 4.083	    & 201    & 0   \\
        & 233809   	& 0.39	 & 4.083	& 0.33	& 6.0	    & 112    & 0   \\
        & 237893   	& 0.46	 & 2.083	& 0.33	& 2.083	    & 2      & 0   \\ 
        & 244019   	& 0.48	 & 2.0		& 0.33	& 4.0	    & 137    & 0   \\
        & 246061   	& 0.34	 & 4.0		& 0.33	& 2.083	    & 142    & 0   \\
        \bottomrule    
        \end{tabular}    
        \caption{\label{tab: thermal_fluctuations_in_RV} Thermal fluctuations in the reverse annealing
        for some failure cases in Table \ref{tab: thermal_fluctuations_in_FW},
        which starts from the minima obtained in Table \ref{tab: thermal_fluctuations_in_FW} at last.}
\end{minipage}
\end{table}
}

\ignore{
\section*{Introducing thermal fluctuations to solve midddle-size problems}

With the limits of the current systems on the effective annealing time ($T_a \lesssim 20\mu s$),
that increasing $T_a$ would not significantly increase the success probability,
we investigate the effect of quantum annealing enhanced by thermal fluctuations
for solving bigger problems, by introducing a pause into the annealing process.
The pause of $T_p=100us|T_a=10us$ length is chosen for factoring integers of 8+8, 9+8 and 10+8 bits,
which was tested, out of $T_p\in \{10, 50, 100\}$, to bring the highest success probability
in finding solutions for the solved cases, factoring 59,989 of 8+8 bits;
and the pause point is chosen from the region $S_p\in[0.33, ..., 0.51]$ of the annealing schedule ($s\in[0,1]$)
until the desired groundstate is found.

As shown by the results in Table \ref{tab: thermal_fluctuations_in_FW},
thermal fluctuations did help in finding groundstates
for factoring some integers of 8+8 bits,
which is impossible for the previous pure annealing with $T_a=10us$,
as well as for factoring integers of bigger size.
However, the success in finding groundstates is not guaranteed.
For the failure cases, we again use the pause of $T_p=100us|T_a=10us$ length 
to reverse annealing from the minima obtained at the latest pause point.
Notet that reverse annealing had been demonstrated have memory of its initial state
and thus can be used as local search.
The revison point of annealing is chosen from $S_p \in [0.51, ..., 0.33]$
for our problem cases.

The results in Table \ref{tab: thermal_fluctuations_in_RV}, 
together with those in Table \ref{tab: thermal_fluctuations_in_FW},
confirm that this zip-up annealing procedure, 
first foward annealing with $S_p=0.33\to0.51$ to obtain a minima
and then from this minima reverse annealing with $S_p=0.51\to0.33$,
enables to converge to groundstates.
This converge depends on the goodness of the minima as the initial state of reverse annealing
and how thermal fluctuations contribute in reverse anenaling.
However, what we can observe from Table \ref{tab: thermal_fluctuations_in_RV} is that 
in general, this pause of $T_p=100us$ length can drive the minimas
roughly in 100+ hamming distance away to groundstates
for the former two problem sizes;
the initial state of higher energy ($\approx 4$)
tends to trigger larger moves for the late reversion points,
compared to that of lower energy ($\approx 2$),
whereas even lager moves require even earlier reversion points.

To evaluate the goodness of the chosen minimas as the initial state of reverse annealing,
we examine their distance towards groundstate 
for some successful cases where groundstate is found.
We found that the chosen minima, that of the minimal energy,
may not guarantee the closest distance toward groundstate,
as shown in Figure \ref{fig: dist_to_gs_4_FW} and Table \ref{tab: thermal_fluctuations_in_FW}.
As a side effect, choosing a minima in very far distance from groundstate, 
like the minima of energy $=2.0$ at $S_p=0.5$ for factoring 49949 of 8+8 bits,
can make the converge to groundstate even harder.
Unless a closer minima towards groundstate can be chosen
after $T_p=100us|T_a=10us$, the minimas obtained by pure annealing $T_a=10us$,
of sufficient energy and in a reasonable distance from groundstate
as shown in Table \ref{tab: tab: minimas_of_pure_FW},
may be a good start for initiating reverse annealing.

As to how thermal fluctuations contribute in reverse anenaling,
we examine different pause lengths for factoring integers of 10+8 bits,
whose groundstates were not found in Table \ref{tab: thermal_fluctuations_in_RV}.
In solving the problem sets, reverse annealing
is also chosen to begin with a different initial states
that are obtained by pure forward annealing $T_a=10\mu s$,
those of little higher energy and in farther distance from groundstate.
We observe that increasing the length of pause may cause larger moves
for late reversion points, whereas the increase of moves
is not obvious for earlier reversion points.
This implies that it may be not feasible to relax 
a minima in a very large distance from groundstate 
by just increasing the length of pause.
Based on these observations, we propose an algorithm for reverse annealing
to solve even bigger problems in the following section.
Note that as the problem size increase,
the scaling of the performance of 
both thermal fluctuations and quantum annealing 
also needs to be considered.
\begin{figure}
    \centering
    \includegraphics[width=.8\textwidth]{figs/updated_results/factoring 10+8 bits/Reverse annealing (Tp|Ta=10us) from the minima obtained from forward annealing (Tp=100us|Ta=10us).png}
    \includegraphics[width=.8\textwidth]{figs/updated_results/factoring 10+8 bits/Reverse annealing (Tp|Ta=10us) from the minima obtained from forward annealing Ta=10us).png}
    \caption{{\label{fig: comparison_of_diff_pause_lengths_and_diff_initial_state}} 
    Comparisons of different pause lengths and initial states for reverse annealing
    for factoring integers of 10+8 bits}
\end{figure}


    

\newpage
}

%% file: solving.tex
\section*{Solving Prime Factorization on D-Wave Advantage 4.1 system}

The results presented in the previous section do not account for the
actual limitations of quantum annealers. In particular, due to
hardware faults, some of the qubits, and some
connections between them are inactive and cannot be tuned during
annealing. These inactive nodes and connections, referred to in the
literature as {\it faulty qubits} and {\it faulty couplings}
respectively, are spread all around the entire architecture,
and are marked in orange in Figures  
 \ref{fig: multiplier_ver2} and \ref{fig: multiplier_ver3} for the
D-Wave Advantage 4.1 annealer, which we have used in
all our experiments in this paper.
Therefore, although it is theoretically possible to create multipliers up to
$21\times 12\text{ bits}$ or $22\times 8\text{ bits}$, these hardware constraints compel us to
test smaller multipliers to avoid faulty qubits and
couplings. An empirical evaluation of possible placements of
multipliers into the Advantage 4.1 system 
leads us to determine an
area of the architecture with no faulty nodes nor couplings that is
suitable for being tested, capable of embedding a multiplier of
maximum size $17\times 8\text{bits}$ with the
configuration of Figures~\ref{fig: multiplier_ver2} and \ref{fig: unify}. All the
experiments in this section will consider these hardware
limitations. 
Also, the experimental evaluation reported in this section was
constrained by the limited amount of QPU time on the Advantage 4.1
annealer we were given access to (600 seconds per month).

\ignore{
\subsection*{The effectiveness of the CFA penalty function}

\ignore{
\begin{figure}[!tbp]
		\centering
		\includegraphics[width=.4\linewidth]{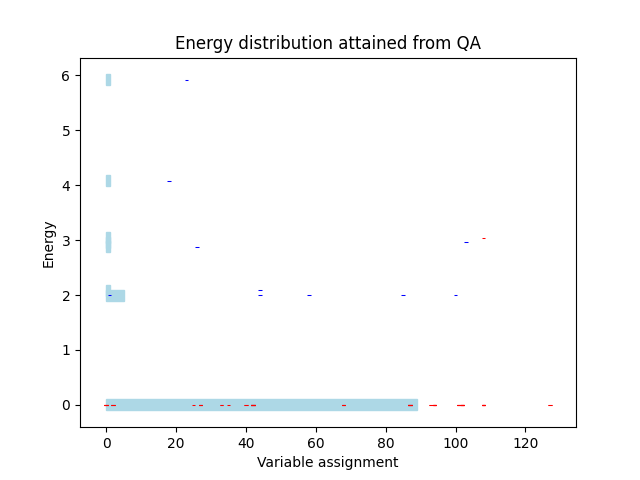}
        \caption{The experimental results
	of the CFA penalty function shown in Figure \ref{fig: CFA0_encoding} on D-Wave Advantage 4.1 via 
	$[T_a = 0.5 \mu s, num\_of\_reads = 100]$}
 \label{fig: CFA0_experiments}
\end{figure}
}

To evaluate the effectiveness of the quantum annealer processes in
reaching the ground state when the CFA penalty function is fed to the
system, we conducted a preliminary experiment using Advantage 4.1. The
penalty function for the CFA described in Figure~\ref{fig: unify} was provided as input to the system and
embedded in one Pegasus tile, with no constraint on the truth value of
both input and output variables. We then run a total of 100
samples and, for each of them, we considered the final energy and the returned value of the input,
output, and carry qubits excluding the ancillae ---i.e. of $in2, in1, enable, c\_in, c\_out, out,
enable\_out$--- to check if they satisfied the CFA or not. 
(Notice that only 16 out of the 128 possible value combinations satisfy
  the CFA.)
}

\subsection*{Initializing qubits}

To factor a specific integer, it is necessary to initialize
several qubits within the multiplier embedding: all qubits associated
with the output bits need to be initialized to represent the target
number for factorization ---e.g., if the output $[P37 ... P00]$ of the $4\times4$-bit
multiplier in Figures~\ref{fig: arith_multiplier} and \ref{fig: multiplier} is forced to
$00100011$ (i.e. $35$), then
the corresponding qubits are initialized respectively to $\set{-1,-1,1,-1,-1,-1,1,1}$; additionally, the variables $c\_in$ and
$in2$ on the most external CFAs should be forced to be 0, as
depicted in Figure \ref{fig: multiplier}, so that their corresponding
qubits should be initialized to $-1$.

D-Wave Advantage interface provides an API, the $fix\_variables()$
function, which allows us to impose desired values on the qubits of
the underlying architecture. This function operates by substituting
the values of the qubits into the penalty function and
subsequently rescaling the resulting penalty function to ensure all
coefficients fall within the limited ranges of biases and
couplings, possibly resulting into a lower $g_{min}$. For instance, if we have the penalty function $\Px = 2 +
4x_1 + x_2 + x_1x_2$ and we set $x_2$ to 1, then the penalty
function becomes $\Pxprime = 2 + 4x_1 + 1 + x_1 = 3 + 5x_1$, which is
then rescaled into $12/5 + 4x_1$ by multiplying it by a $4/5$ factor
in order to fit the bias of $x_1$ into the $[-4,4]$ range, thus
reducing $g_{min}$ by multiplying it the same  $4/5$ factor.
On the one hand, this substitution simplifies the penalty function
by removing one binary variable; on the other hand, it can
hurt the minimal gap due to coefficient
rescaling.

To cope with the latter problem, we propose an alternative method to initialize qubits on a quantum device. We can partially
influence the quantum annealer to set a specific truth value
for a qubit by configuring {\it flux biases}
\cite{flux-bias}. In particular, if we want to impose the value $s_i\in\{-1,1\}$ on a qubit, 
we set the flux bias for that qubit as 
$\phi_i = 1000\phi_0s_i$, where $\phi_0$ is
the default annealing flux-bias unit of the DWave system 4.1,
whereas $1000$ is an empirical value we choose based on our experience.

The experiments suggested a further minor improvement in the CFA
encoding.
Since there may be more than one penalty function with the optimum
value of  $g_{min}$, we make a second call to an OMT solver in which
we fix $g_{min}$ and 
ask the solver to find a solution which also minimizes the number of
those falsifying assignments which
make the penalty function equal to $g_{min}$.
The intuition here is to minimize the possibility of the annealer to
get excited from ground states to first excited un-satisfying states.
(Hereafter we refer as
``CFA1'' the CFA encoding obtained with this improvement and as ``CFA0''
the basic one.)

In Table \ref{tab: foward_annealing_345} we compare the performances
of the two initialization techniques on small prime factorization
problems, with the annealing time $T_a$ set to $10\mu s$. The column labeled $\#(P_F = 0)$ reports how many occurrences of 0-energy samples are
obtained out of 1000 samples. We noticed that flux biases (with CFA1) outperform the native API, having a higher probability of reaching the global minimum. All the experiments from now on assume qubit initialization is done by tuning flux biases.

\ignore{
\begin{tabular}{|l|c|r|r|}
            \toprule
           \multirow{2}{*}{size}     & \multirow{2}{*}{inputs}  & \multicolumn{2}{c}{CFA0($\#(P_F = 0)$)} & \multicolumn{2}{c}{CFA1($\#(P_F = 0)$)}\\
                                     &                          &  api & flux-bias                        & api & flux-bias \\
           \midrule
            \multirow{3}{*}{3$\times$3}
            & 25(5$\times$5)  &	{\bf 161}   & 308		      & 327  & 136   \\
            & 35(5$\times$7)  &	389         & 711		      & 410  & {\bf 951}   \\
            & 49(7$\times$7)  &	450         & 906		      & 344  & {\bf 997}   \\
            \midrule
            \multirow{3}{*}{4$\times$4}
            & 121(11$\times$11)   & {\bf 17}  & 63    		& 9    & 0     \\
            & 143(11$\times$13)   & 40        & 129	  	  & 122  & {\bf 67}    \\
            & 169(13$\times$13)   & {\bf 31}  & 312   		& 84   & 5     \\
            \midrule
            \multirow{15}{*}{5$\times$5}
            & 289(17$\times$17)   & {\bf 5}   &	1  		& 3    &  0     \\
            & 323(17$\times$19)   & {\bf 2}   &	7  		& 22   &  0     \\
            & 361(19$\times$19)   & 1         &	1  		& 11   &  {\bf 3}     \\
            & 391(17$\times$23)   & 6         &	119		& 5    &  {\bf 9}     \\
            & 437(19$\times$23)   & {\bf 17}  &	67 		& 3    &  0     \\
            & 493(17$\times$29)   & {\bf 3}   &	4  		& 8    &  2     \\
            & 527(17$\times$31)   & 21        &	91 		& 6    &  {\bf 37}    \\
            & 529(23$\times$23)   & 5         &	8  		& 0    &  {\bf 8}     \\
            & 551(19$\times$29)   & 0         &	24 		& 2    &  {\bf 4}     \\
            & 589(19$\times$31)   & 16        &	7  		& 1    &  {\bf 52}    \\
            & 667(23$\times$29)   & 0         &	3  		& 8    &  {\bf 105}   \\
            & 713(23$\times$31)   & 11        &	26 		& 2    &  {\bf 138}   \\
            & 841(29$\times$29)   & 5         &	148		& 14   &  {\bf 7}     \\
            & 899(29$\times$31)   & 17        &	222		& 7    &  {\bf 343}   \\
            & 961(31$\times$31)   & 1         &	37 		& 1    & 338  \\ 
            \bottomrule
        \end{tabular}
}

\begin{table}[t]
  \begin{minipage}[b]{0.4\linewidth}  
    \centering
    \scriptsize
    \begin{tabular}{|l|l|r|r|}
            \toprule
           \multirow{2}{*}{Size}     & \multirow{2}{*}{Input $N$}  & CFA0 & CFA1\\
                                     &                          &  $\#(P_F = 0)$ &  $\#(P_F = 0)$  \\
           \midrule
            \multirow{3}{*}{3$\times$3}
            & 25 (5$\times$5)  &	{\bf 161}   & 136   \\
            & 35 (5$\times$7)  &	389         & {\bf 951}   \\
            & 49 (7$\times$7)  &	450          & {\bf 997}   \\
            \midrule
            \multirow{3}{*}{4$\times$4}
            & 121 (11$\times$11)   & {\bf 17}  & 0     \\
            & 143 (11$\times$13)   & 40        & {\bf 67}    \\
            & 169 (13$\times$13)   & {\bf 31}    & 5     \\
            \midrule
            \multirow{15}{*}{5$\times$5}
            & 289 (17$\times$17)   & {\bf 5}    &  0     \\
            & 323 (17$\times$19)   & {\bf 2}   &  0     \\
            & 361 (19$\times$19)   & 1          &  {\bf 3}     \\
            & 391 (17$\times$23)   & 6             &  {\bf 9}     \\
            & 437 (19$\times$23)   & {\bf 17}     &  0     \\
            & 493 (17$\times$29)   & {\bf 3}       &  2     \\
            & 527 (17$\times$31)   & 21            &  {\bf 37}    \\
            & 529 (23$\times$23)   & 5            &  {\bf 8}     \\
            & 551 (19$\times$29)   & 0             &  {\bf 4}     \\
            & 589 (19$\times$31)   & 16            &  {\bf 52}    \\
            & 667 (23$\times$29)   & 0             &  {\bf 105}   \\
            & 713 (23$\times$31)   & 11          &  {\bf 138}   \\
            & 841 (29$\times$29)   & 5            &  {\bf 7}     \\
            & 899 (29$\times$31)   & 17           &  {\bf 343}   \\
            & 961 (31$\times$31)   & 1            & {\bf 338}  \\ 
            \bottomrule
        \end{tabular}
    \captionsetup{labelformat=parens,labelsep=space}
    \subcaption{\label{tab: foward_annealing_345} Comparison of the two initialization techniques on prime factorization of small numbers, with $T_a = 10 \mu s$.}
  \end{minipage}
  \hfill
  \begin{minipage}[b]{0.58\linewidth}  
    \centering
    \scriptsize
    \begin{tabular}{|l|l|r|c|l|l|r|}
            \toprule
           \multirow{1}{*}{Size}     & \multirow{1}{*}{Input $N$}  & $\#(P_F = 0)$ & & \multirow{1}{*}{Size}     & \multirow{1}{*}{Input $N$}  & $\#(P_F = 0)$\\
            \midrule
            \multirow{10}{*}{7$\times$7}
            & 10,033 (127$\times$79)	   & 0         & & \multirow{10}{*}{8$\times$8} & 49,447 (251$\times$197)	  & 0    \\
            & 10,541 (127$\times$83)	   & {\bf 1}   & &                              & 49,949 (251$\times$199)	  & 0    \\
            & 11,303 (127$\times$89)	   & 0         & &                              & 52,961 (251$\times$211)	  & 0    \\
            & 12,319 (127$\times$97)	   & 0         & &                              & 55,973 (251$\times$223)	  & 0    \\
            & 12,827 (127$\times$101)	   & {\bf 1}   & &                              & 56,977 (251$\times$227)	  & 0     \\
            & 13,081 (127$\times$103)	   & {\bf 2}   & &                              & 57,479 (251$\times$229)	  & 0    \\
            & 13,589 (127$\times$107)	   & {\bf 10}  & &                              & 58,483 (251$\times$233)	  & 0    \\
            & 13,843 (127$\times$109)	   & 0         & &                              & 59,989 (251$\times$239)	  & {\bf 2}    \\
            & 14,351 (127$\times$113)	   & 0         & &                              & 60,491 (251$\times$241)	  & 0      \\
            & 16,129 (127$\times$127)	   & {\bf 7}   & &                              & 63,001 (251$\times$251)	  & 0     \\
            \bottomrule
        \end{tabular}
    \captionsetup{labelformat=parens,labelsep=space}
    \subcaption{\label{tab: foward_annealing_78} Prime factorization of the 10 biggest $7\times7$ and $8\times8$ numbers configuring flux biases, with $T_a=10\mu s$. }
  \end{minipage}
  \caption{Results of standard forward annealing to solve prime factorization.}  
\end{table}

\ignore{
To cope with the latter problem, we propose three alternative methods to initialize qubits on a quantum device:
\begin{itemize}
	\item {\it Extra chaining}: for each variable that we want to
          initialize we choose an unused qubit directly connected to
          it with an active coupling and use it to impose its truth
          value through qubit chaining. In detail, if we want to force
          qubit $x$ to be true (resp. false), we consider the neighbor
          unused qubit $x'$ and extend the multiplier penalty function
          $\Px$ with an equivalence chain $2 - 2xx'$. Then, we call
          $fix\_variables()$ so that $x'$ is set to true
          (resp. false), meaning that each occurrence of $x'$ is
          replaced by +1 (resp. -1). This way the simplification of
          $x'$ results in a variation of the offset and of the bias
          of $x$ only, so there is no risk of rescaling.
          
	\item {\it Ad-hoc encoding for the CFAs (custom CFAs)}: we
          re-encode CFAs that require some qubits to have a specific
          truth value through OMT solving, by adding Boolean
          constraints to fix the value of some qubits. Consequently,
          we create a library of ad-hoc CFA penalty functions for all
          possible input values. For instance, suppose we want to set
          the value of $c\_{in}$ to false. Then the OMT solver is fed the
          extended formula
          $F'(\xs) = F(\xs) \wedge \neg c\_{in}$ to
          generate a new penalty function, forcing the placement of
          qubits to follow the schema of Figure \ref{fig: unify}. Notice
          that this re-encoding should still guarantee all
          combinations of input values do not make the resulting
          biases and couplings of the penalty function out of the
          limited range due to sharing, thus all arithmetical
          constraints introduced by qubit sharing and virtual chaining
          are also considered by the OMT solver. This re-encoding
          process produces penalty functions with increased minimal
          gaps, $g_{min} \in [3, 18]$, theoretically increasing the
          probability of the QA to reach a global minimum. We remark
          that using custom CFAs to embed a multiplier initialized to
          a specific number does not affect performance, since we
          need to generate the set of CFAs offline only once. To
          generate a custom multiplier we then extract the various
          CFAs from the pre-computed library, with no computational
          overhead.
	\item {\it Altering flux biases}:
	instead of modifying the penalty functions, we can partially
        influence the quantum annealer to set a specific truth value
        for a qubit by configuring flux biases. \JDSIDENOTE{cite something here?}
        In particular, we set
        the flux biases of qubits associated with fixed variables by
        $\phi_i\defas \alpha s_i$, 
	where $s_i \in \{-1, 1\}$ is the spin value to impose if the variable is false or true, respectively,
	and $\alpha$ is the strength of the flux bias.
	We empirically set $\alpha\defas 1000\phi_0$ for all problem
        instances, where $\phi_0\defas 1.4303846404537006e-05$ is the default value in D-Wave Advantage systems. 
\end{itemize}
A comparison of the four initialization approaches 
for QA solving small prime factorization problems is shown in
Figure~\ref{tab: diff_initializations}.
Among the four approaches considered, tuning the flux biases to
influence the spin of the qubits demonstrates superior performance,
increasing the probability of finding ground states.
Nevertheless, it is
important to notice that this approach directly modifies the
configurations of the quantum annealer. Investigating the impact of
these configurations on the solving process fall outside the scope of
this study, as our focus is solely on providing encoding for prime
factorization without altering the properties of the quantum
systems. Consequently, a more in-depth analysis of the influence of
flux bias on solving prime factorization is left for future research
directions.
Regarding the performance enhancement achieved through extra chains
and ad-hoc CFA encoding, the results indicate that their impact is not
much significant. Although there is a clear effect of these techniques in solving $3\times 3$ bit problems, their impact on $4\times 4$, $5\times 5$, and $6\times 6$ bit prime factorizations shows minimal improvement compared to the default approach. Therefore, for the remainder of this work, we will assume that the initialization is performed using the default procedure of the $fix\_variable()$ function. 

\begin{figure}[!tbp]
	\begin{minipage}{.5\linewidth}
		\includegraphics[width=1\linewidth]{figs/results/multiplier_ver2_CFA0/diff_initializations/Initialize the multiplier Hamiltonian with API VS ad hoc CFA Hamiltonians VS extra chains VS external flux biasesrange(3, 6).png}
	\end{minipage}
	\begin{minipage}{.5\linewidth}
		\includegraphics[width=1\linewidth]{figs/results/multiplier_ver2_CFA0/diff_initializations/Initialize the multiplier Hamiltonian with API VS ad hoc CFA Hamiltonians VS extra chains VS external flux biasesrange(6, 7).png}
	\end{minipage}
	\caption{\label{tab: diff_initializations} Factoring $3\times 3, 4\times 4, 5\times 5$ bit and $6\times 6$ bit integers with $[num\_of\_reads = 1000]$
	and with $[T_a = 1\mu s]$, $[T_a = 10\mu s]$, $[T_a = 20\mu s]$ and $[T_a = 20\mu s]$ respectively, 
	in different ways of initialization. CHANGE LEGEND
      }
\end{figure}
}

\ignore{
\subsection*{Exploiting Boolean Value Propagation}

Table \ref{tab: PF_results_with_API} shows the detailed results of factoring integers from 6 up to 12 bits by initializing variables through the D-Wave API ($fix\_variables()$),
with 1000 samples and $T_a \leq 20\mu s$. We can see how our
encoding for multipliers works well up to $5\times5\text{ bits}$
integers, whereas for larger numbers we rarely reach the ground state.

To overcome in part this issue, we can produce a simpler penalty 
function \Pxa{} by simplifying upfront the input Boolean formula $\Fx$.
  As we have noticed above, some qubits are initialized to
  some fixed values, corresponding to Boolean values of variables in
  the Boolean formula $F(\allx)$ representing the multiplier. It may be the case, however, that
  some such values force the value of some other Boolean variables in $F(\allx)$. (E.g., consider
  $F(\allx)\defas ... \wedge (x_1\iff (x_2 \vee x_3))$: if $x_1$ is
  forced to be $\bot$, then both $x_2$ and $x_3$ are forced to be $\bot$; if
  instead $x_2$ is forced to be $\top$, then $x_1$ is forced to be
  $\top$.)
  Thus, we can detect these derived values by applying {\it
    Boolean value propagation (BVP)} to \Fx. Then we set the
  $\set{-1,1}$ values of  the corresponding qubits by calling
  $fix\_variables()$. This allows us to further reduce the number of binary
  variables for the annealer to cope with.

\ignore{Suppose we have a CNF propositional formula and a set of initial literals \textbf{\underline{l}}. For each literal $\ell$ in \textbf{\underline{l}} we remove it from the set and propagate it into the formula, i.e. we remove $\neg \ell$ from every clause that contains $\neg \ell$ and remove each clause containing $\ell$. If during this simplification a new unit clause is generated, then we add the literal in this unit clause to \textbf{\underline{l}}. We repeat the procedure until no more literals are in \textbf{\underline{l}}. \\
For example, let a given formula be 
$x_1 \wedge (\neg x_1 \vee x_2) \wedge (x_1 \vee x_3) \wedge (\neg x_2 \vee x_4)$. Being a unit clause, $x_1$ is part of \textbf{\underline{$\ell$}}.
Through the unit propagation on $x_1$,
we remove $\neg x_1$ from the second clause
and the third clause entirely,
obtaining $x_2 \wedge (\neg x_2 \vee x_4)$.
Furthermore, the resulting formula can be simplified further
by applying BVP to $x_2$,
finally reducing the original formula to $x_4$. \\
In the context of integer factorization, we adopt the bit-vector
representation of the output number as the initial set of literals to
be propagated by Boolean Constraint Propagation (BVP).
We then set the truth value of all the unit propagated variables by
calling $fix\_variables()$.
}
The results in the last column of Table \ref{tab: PF_results_with_API}
shows the improvements achieved by incorporating BVP as a
pre-processing step, indicating the number of propagated values
and the new count of samples reaching the ground state, up until
$6\times 6$ multipliers.
\ignore{We notice how the number of propagated
values only grow by 3 every time we increase the input number size
by one, thus the effects of BVP are less evident with a prime
factorization of larger numbers.} We notice that although
the number of propagated values grows only linearly with the bit-size
of the factors, we have a significant increase in the number of
solutions the annealer can find.
Notice that, since modern Boolean reasoning tools 
are capable of
performing BVP very efficiently, the overhead of performing BVP as a
preprocessing step 
is negligible.
}


\subsection*{Exploiting thermal relaxation}

  In order to test the limits of 
  the flux-bias initialization,
  we applied it to factoring the 10 largest numbers
  of $7\times7$ and $8\times8$ bits with the same annealing time as the previous experiments ($T_a=10\mu s$.)
    The results, reported in Table \ref{tab:
      foward_annealing_78}, suggest that the success probability of getting a solution for 16-bit numbers is almost null. Increasing the annealing time $T_a$, however, would probably not significantly increase the success probability; to further improve the solving performances, we investigate the effectiveness of {\it thermal relaxation} \cite{Marshall2019}
  on solving our problems.
  This technique is integrated into the DWave system by introducing a {\it pause} $T_p$
  at a specific point $S_p$ during the annealing process, with $S_p\in[0,1]$.
  We tested it to solve $8\times 8$, $9\times 8$ and $10\times 8$-bit
  factorization problems. 

\begin{table}[t]
  \begin{minipage}[b]{1\linewidth}  
    \centering
    \scriptsize
    \begin{tabular}[pos]{|c|c|c|c|r|c|c|c|c|c|r|c|c|c|c|c|r|}
            \toprule
            \rot{90}{Size} & \rot{90}{Input $N$} & \rot{90}{$S_p$} & \rot{90}{$min(P_F)$} & \rot{90}{\#$(P_F=0)$} & & \rot{90}{Size} & \rot{90}{Input $N$} & \rot{90}{$S_p$} & \rot{90}{$min(P_F)$} & \rot{90}{\#$(P_F=0)$} & &
            \rot{90}{Size} & \rot{90}{Input $N$} & \rot{90}{$S_p$} & \rot{90}{$min(P_F)$} & \rot{90}{\#$(P_F=0)$}\\
            \midrule
            \multirow{10}{*}{\rot{90}{8$\times$8}} & 49,447 (251$\times$197)     & 0.38	    & 0.000	    & \textbf{1}  & &\multirow{10}{*}{\rot{90}{9$\times$8}} & 100,273 (509$\times$197) 	  & --	    & 4.083	    & 0           & &\multirow{10}{*}{\rot{90}{10$\times$8}}  & 201,137 (1021$\times$197) 	& --	    & 6.167	    & 0    \\
                                                   & 49,949 (251$\times$199)     & --	      & 4.083	    & 0           & &                                       & 101,291 (509$\times$199)	  & --	    & 8.083	    & 0           & &                                         & 203,179 (1021$\times$199)	& --	    & 8.000	    & 0        \\
                                                   & 52,961 (251$\times$211)     & --	      & 6.000	    & 0           & &                                       & 107,399 (509$\times$211)	  & --	    & 4.000	    & 0           & &                                         & 215,431 (1021$\times$211)	& --	    & 6.083	    & 0   \\
                                                   & 55,973 (251$\times$223)     & 0.33	    & 0.000	    & {\bf 6}     & &                                       & 113,507 (509$\times$223)	  & --	    & 8.083	    & 0           & &                                         & 227,683 (1021$\times$223)	& 0.34	  & 0.000	    & \textbf{1}  \\
                                                   & 56,977 (251$\times$227)     & 0.33	    & 0.000	    & \textbf{1}  & &                                       & 115,543 (509$\times$227)	  & 0.33	  & 0.000	    & \textbf{1}  & &                                         & 231,767 (1021$\times$227)	& --	    & 8.083	    & 0  \\
                                                   & 57,479 (251$\times$229)     & 0.33	    & 0.000	    & \textbf{3}  & &                                       & 116,561 (509$\times$229)	  & --	    & 6.000	    & 0           & &                                         & 233,809 (1021$\times$229)	& --	    & 8.000	    & 0   \\
                                                   & 58,483 (251$\times$233)     & --	      & 6.083	    & 0           & &                                       & 118,597 (509$\times$233)	  & --	    & 4.000	    & 0           & &                                         & 237,893 (1021$\times$233)	& --	    & 6.000	    & 0 \\
                                                   & 59,989 (251$\times$239)     & 0.33	    & 0.000	    & \textbf{43} & &                                       & 121,651 (509$\times$239)	  & 0.33	  & 0.000	    & \textbf{1}  & &                                         & 244,019 (1021$\times$239)	& --	    & 6.250	    & 0  \\
                                                   & 60,491 (251$\times$241)     & 0.38	    & 0.000	    & \textbf{1}  & &                                       & 122,669 (509$\times$241)	  & --	    & 8.167	    & 0           & &                                         & 246,061 (1021$\times$241)	& --	    & 6.167	    & 0 \\
                                                   & 63,001 (251$\times$251)     & --	      & 2.000	    & 0           & &                                       & 127,759 (509$\times$251)	  & 0.36	  & 0.000	    & \textbf{1}  & &                                         & 256,271 (1021$\times$251)	& 0.35	  & 0.000	    & \textbf{2}\\
            \bottomrule
        \end{tabular}
    \captionsetup{labelformat=parens,labelsep=space}
    \subcaption{\label{tab: thermal_fluctuations_in_FW} Prime factorization of $8\times8$, $9\times8$ and $10\times8$-bit numbers, with $T_a = 10\mu s$ and pause $T_p = 100\mu s$.}
  \end{minipage}
  \begin{minipage}[b]{1\linewidth}  
    \centering
    \scriptsize
    \begin{tabular}{|l|c|rr|crrr|}
        \toprule
        \multirow{2}{*}{Size}   & \multirow{2}{*}{Input $N$}  & \multicolumn{2}{c}{Forward annealing} & \multicolumn{4}{c}{Reverse annealing}\\
                                &                          & $S_p$ & $min(P_F)$                      & $S'_p$ & $P_F$ & $\Delta HAM$ & \#$(P_F=0)$ \\
        \midrule
        \multirow{4}{*}{8$\times$8}
        & 49,949 (251$\times$199)     & 0.50	 & 2.000    & 0.33	& 0.000	& 233	& \textbf{7}     \\
        & 52,961 (251$\times$211)     & 0.35	 & 2.000    & 0.41	& 0.000	& 177	& \textbf{1}     \\
        & 58,483 (251$\times$233)     & 0.33	 & 2.083    & --	& 4.000	& 144   & 0      \\
        & 63,001 (251$\times$251)     & 0.51	 & 2.000    & 0.35	& 0.000	& 168	& \textbf{4}     \\
        \midrule
        \multirow{7}{*}{9$\times$8}
        & 100,273 (509$\times$197)	  & 0.36	 & 4.000		& --	& 4.083	& 198   & 0     \\
        & 101,291 (509$\times$199)	  & 0.44	 & 4.000		& --	& 4.000	& 2     & 0     \\
        & 107,399 (509$\times$211)    & 0.51	 & 4.000    & --	& 4.000	& 79    & 0      \\
        & 113,507 (509$\times$223)    & 0.38	 & 2.000    & --	& 2.000	& 71    & 0      \\
        & 116,561 (509$\times$229)    & 0.36	 & 4.000	    & 0.37	& 0.000	& 98	& \textbf{35}    \\
        & 118,597 (509$\times$233)    & 0.33	 & 2.000	    & --	& 4.000	& 201   & 0      \\
        & 122,669 (509$\times$241)    & 0.48	 & 4.083	& 0.36	& 0.000	& 129	& \textbf{7}     \\
        \midrule 
        \multirow{8}{*}{10$\times$8}
        & 201,137 (1021$\times$197)	    & 0.38	 & 2.000		& --	& 2.000		& 6      & 0   \\
        & 203,179 (1021$\times$199)	    & 0.34	 & 4.000		& --	& 4.083	    & 218    & 0   \\
        & 215,431 (1021$\times$211)	    & 0.4	   & 4.000		& --	& 4.000		& 228    & 0   \\
        & 231,767 (1021$\times$227)   	& 0.33	 & 2.083	& --	& 4.083	    & 201    & 0   \\
        & 233,809 (1021$\times$229)   	& 0.39	 & 4.083	& --	& 6.000	    & 112    & 0   \\
        & 237,893 (1021$\times$233)   	& 0.46	 & 2.083	& --	& 2.083	    & 2      & 0   \\ 
        & 244,019 (1021$\times$239)   	& 0.48	 & 2.000		& --	& 4.000	    & 137    & 0   \\
        & 246,061 (1021$\times$241)   	& 0.34	 & 4.000		& --	& 2.083	    & 142    & 0   \\
        \bottomrule    
        \end{tabular} 
    \captionsetup{labelformat=parens,labelsep=space}
    \subcaption{\label{tab: thermal_fluctuations_in_RV} Results of performing reverse annealing on the problem instances not solved in Table \ref{tab: thermal_fluctuations_in_FW}, with $T_a = 10\mu s$ and $T_p = 10\mu s$. The label $\Delta HAM$ reports the Hamming distance between the forward annealing lowest energy sample and the reverse annealing lowest energy sample.}
  \end{minipage}
  \caption{Results about prime factorization solved through QA, exploiting thermal relaxation.}  
\end{table}

  In the experiments, the pausing time $T_p$ was set to $100 \mu s$,
  whereas the pause point $S_p$ is selected in the set $\{0.33, 0.34, ...,
  0.51\}$ and tested in ascending order until the ground state is found.
The results illustrated in Table \ref{tab: thermal_fluctuations_in_FW},
if compared with
  these in Table \ref{tab: foward_annealing_78}, indicate the positive impact of thermal relaxation. 
Ground states were successfully reached for some 18-bit numbers (the largest being 256271), although challenges persist with most numbers of that size.

\subsection*{Exploiting quantum local search}
  For the factorization problems in Table \ref{tab: thermal_fluctuations_in_FW}
  that did not end up in the global minimum,
  we further exploited {\it quantum local search}, consisting of refining a sub-optimal state to reach the global minimum.
  Quantum local search is implemented in the DWave system by mean of {\it reverse annealing} (RV)\cite{reverse2017}. The annealer is 
  initialized in a local minimum, whereas the annealing process
  starts from $s=1$ moving towards $s'=0$ and then returning back to $s=1$.
  We remark that reverse annealing admits pauses during the process: in this case, the system pauses for $T_p$ microseconds at a middle point $s'=S'_p$. 

  In our experiments, we chose the lowest-energy state from table
  \ref{tab: thermal_fluctuations_in_FW} as the initial state of RV. If
  multiple lowest-energy samples are obtained with different $S_p$
  values, we pick the one whose pause is performed later. 
  The pause points for RV were tested in decreasing order (in opposition to forward annealing when we opted for the ascending order) until a ground state was found. The results are reported in Table \ref{tab: thermal_fluctuations_in_RV}. 
  We observe that reverse annealing, enhanced by thermal relaxation,
  helps in solving up to $9\times 8$-bit factorization problems. We
  also reported the Hamming distance $\Delta HAM$ between the
  lowest-energy state from forward and reverse annealing, showing how
  much a sample moved from one minimum to another, possibly a
  ground state. 
  
  For the instances that still failed to reach a solution, we investigated the impact of different pause lengths
  for RV to find ground states.
  The main observation from this additional analysis is that, given a low-energy initial state: ($i$)
  increasing the pause length and performing the pause at a late annealing point can help reverse annealing in jumping larger Hamming distances; 
  ($ii$) increasing the pause length and triggering the pause at early annealing points
  cannot make RV move \JSChange{even farther}.
  From these observations, we could imply that if the initial state of a reverse annealing process is \JSChange{very} far from 
  the ground state, it could be hard to reach the global minimum by only increasing 
  the pause length.
  \JSChange{However, the local minimum used for the initial state of RV, 
  which is obtained by standard annealing,}
  tends to be highly excited (i.e., with high energy and very far from the ground state), as the problem size increases.
  
  In the next section, we \JSChange{follow the {\it iterated reverse
      annealing\cite{PhysRevA.100.052321}} approach, 
  which was studied numerically in a closed-system setting},
   and propose an iterative strategy 
  for the DWave system to solve bigger problems. 
  \JSChange{The} goal is to converge to a low-energy state 
  that can be \JSChange{used as the initial state for single-iteration RV}
  to reach the global minimum 
  with an \JSChange{effective} pause $T_p$.

\ignore{
The results of this two-step search, 
when used in combination with the three soft initialization techniques, are shown in Figure \ref{fig: reverse}. For different output numbers, we tested 6 configurations: 3 based on a combination of initialization techniques ($fix\_variables()$ + unit propagation, custom CFAs, and extra chaining) with no reverse annealing, and 3 with the same initialization techniques discussed before enhanced by reverse annealing. The upper plot reports the pausing point for each factorization instance, obtained empirically by testing several values and reporting the one obtaining lower energy samples; the bottom plot shows the lowest energy sample for each configuration. 
Upon analysis, it is evident that the three initialization techniques discussed in the previous subsections do not show a consistent improvement in efficiency when utilized in conjunction with the two-step quantum search. Additionally, certain techniques fail to benefit from the application of reverse annealing, particularly when employing Boolean Value Propagation before fixing the initial qubit values using the D-Wave APIs. Consequently, we exclusively employ the D-Wave API $fix\_variables()$ for qubit initialization when employing reverse annealing.} 

%
\ignore{
 From our experiments, reverse annealing contributes significantly in
 two key aspects: $i$) at least one sample achieves a ground state with
 an energy of 0 for each problem, ensuring all variables, both
 ancillary and non-ancillary ones are correctly assigned; $ii$) a
 a considerable number of samples obtained from reverse annealing
 correctly factorize the output prime number, surpassing the results
 obtained solely through forward annealing. 
 }

%% file: IRV_improvement.tex
\subsection*{Solving prime factorization with iterated reverse annealing (IRV)}


\ignore{
To understand the reason why RV succeeds in solving those problem instances in Table \ref{tab: large_factoring}
but fails to solve other problem instances we tested,
we examine how far the minimas that are obtained by the pause-enhanced FW
away from the desired ground states.
We observe that for those success cases, the distances between them,
in terms of qubit flips, are roughly 20,
or in terms of CFA flips, are less than 2.
This closeness of the minima to the desired ground state is assumed to provide
a good starting point for RV, whereas the minima quite far from the desired ground state
is assumed hard for RV to reach the ground state,
even a long pause is introduced in RV to drive large thermal fluctuation.
The latter hard situation is further explored in this section via iterated RV (IRV),
to answer the question whether the groundstate far away from currently obtained minima
can be reached by RV iteratively.
}

In general, we assume that starting reverse annealing from a state 
\JSChange{that is close to the ground state}
could be beneficial in \JSChange{finding the solution.}
We remark, however, that we have no prior knowledge of the solution. 
To cope with this missing information, 
we assumed that a low-energy state may be closer to the ground state 
and our proposal is built on top of this assumption.

The IRV strategy starts by running a standard forward annealing process, 
with thermal relaxation disabled. 
The obtained lowest-energy state is selected as 
the starting point for the subsequent iterations of the algorithm. 
At each iteration of the IRV, we execute a batch of RV processes,
with several pause lengths $T_p$ and pausing points $S_p$ taken into consideration, until we obtain a lower-energy space. The {\it lower-energy space} refers to the set of lower-energy states retrieved in one iteration whose energy is {\it below} the starting point.
\JSChange{Once that space has been retrieved,}
we check if \JSChange{there is a ground state in that space:} 
when this happens, we have \JSChange{the} solution for the problem and we stop the entire procedure;
\JSChange{otherwise, } this procedure is iterated until the system 
finds the ground state or hits a certain number of iterations.

It is not trivial to determine how long a pause should be and when to trigger it
for the intermediate iterations to gradually approach the ground state.
	Based on the \JSChange{previous} observations, 
	we chose a set of pause lengths 
	e.g., $\{1, 10, 30, 50, 100\}\mu s$
	and a set of pause point, e.g., $\{0.46, ..., 0.33\}$,
	  adapting those parameters 
	\JSChange{to the initial states of this iteration.}
We tested IRV on the DWave Advantage System 4.1 by trying to factorize the numbers 1,027,343, 4,111,631, and 16,445,771 using respectively a 12$\times$8, 14$\times$8, and 16$\times$8-bit multiplier. 
    The experiments consider the assumptions discussed in the previous paragraphs, a further analysis of these conditions is left as future work.
	Table \ref{tab: IRV} reports the successful search paths of IRV in finding the ground state,
	demonstrating that IRV is effective in reaching a solution even 
	from an excited state very far away from the minimum, by approaching it gradually.
	We highlight that from our experiments it was impossible for standard reverse annealing to factor 4,111,631 even with a $600\mu s$ pause.

We also propose a variant of the IRV strategy discussed above. 
From the failed factorization of 16,445,771, 
we noticed that the last iteration got stuck in the local minimum 
even with a pause of $100\mu s$. 
To cope with \JSChange{this issue, }
we opted to focus on triggering \JSChange{long distances.}
\JSChange{This is done}
by increasing the pause length at each iteration, i.e., $T_p\in\{100, 200\}\mu s$.
\JSChange{Correspondingly,}	
we simplify the choice of the starting state for an iteration,
 \JSChange{choosing} the lowest-energy state 
 \JSChange{as the initial state of each iteration.}
 The experimental results shown in Table \ref{tab: more_effient_IRV}
	demonstrate the improvement of this variant of IRV,
	in terms of fewer iterations required to reach the solution, 
	at the cost of more QPU time. 
	\JSChange{Notice that in the case of the 23-bit number, 8,219,999,
	we use a pause of $1\mu s$. This is
	due to the fact the initial state is highly excited
	and a $1 \mu s$ pause can still trigger a relatively long distance, saving QPU time.}
	\JSChange{According to the results in Table \ref{tab: IRV},}
	we highlight 
	how the fourth iteration highly benefits from the long pause. 
	Despite starting from a local minimum 
	\JSChange{that is very far away from the solution,} 
	\JSChange{the long pause enables RV to}
	travel long Hamming distances
	\JSChange{and reach a local minimum closer to our solution}.
	\JSChange{This closer state provides a good initial state for the last-iteration RV 
	to find the solution successfully.}

\begin{table}[t]
  \begin{minipage}[b]{1\linewidth}  
    \centering
    \scriptsize
    \begin{tabular}{|c|c|c|c|c|rr|c|}
	\toprule
	Size & Input $N$ & \# & $T_p$ & $S_p$ & $min(P_F)$ & $min(P_F)_{new}$  & $(HAM, \Delta HAM, HAM_{new})$ \\
	\midrule
	 & $1,027,343$  & 1 & 1 & 0.31 & 10.167 & 4.000 & (263, 142, 151)\\
	$12\times 8$ & ($4093\times 251$)	& 2 & 1 & 0.38 & 10.167 & 4.083 & (128, 122, 58)\\
	&	& 3 & 100 & 0.38 & 4.083 & 0[\textbf{2}] & (58, 58, 0)\\
	\midrule
	 &  & 1 & 1 & 0.35 & 18.167 & 8.083 & (290, 353, 249) \\
	$14\times 8$ & $4,111,631$	& 2 & 1 & 035 & 16.000 & 6.000 			& (273, 219, 240) \\
	& $(16381\times 251)$	& 3 & 50 & 0.37 & 10.000 & 2.083 	& (277, 280, 85)\\
	&	& 4 & 10 & 0.38 & 2.083 & 0[\textbf{67}]  & (85, 85, 0)\\
	\midrule
	 & & 1 & 1 & 0.34 & 18.333 & 6.083 & (, 294,) \\
	$16\times 8$ & 16,445,771	& 2 & 10 & 0.35 & 10.000 & 4.000 		& (, 374, )\\
	& $(251 \times 65521)$	& 3 & 50 & 0.36 & 6.083 & 4.167 	& (, 292, )\\
	&	& 4 & 100 & 0.39 & 4.167 & 4.000		& (, 8,)\\
	\bottomrule
	\end{tabular}
     \captionsetup{labelformat=parens,labelsep=space}
    \subcaption{\label{tab: IRV} Results of the original IRV algorithm. For the 16$\times$8 problem, since no ground state has been retrieved, no comparison of Hamming distances with the ground state are provided (thus $HAM$ and $HAM_{new}$ are left empty).}
  \end{minipage}
  \begin{minipage}[b]{1\linewidth}  
    \centering
    \scriptsize
    \begin{tabular}{|l|c|c|c|c|rr|c|}
	\toprule
	Size & Input $N$ & \# & $T_p$ & $S_p$ & $min(P_F)$ &   $min(P_F)_{new}$ & $(HAM, \Delta HAM, HAM_{new})$ \\
	\midrule
	$13\times 8$ & $2,055,941$ & 1 & 100 & 0.38 & 14.083 & 6.083 & (204, 185, 55) \\
	& $(8191\times 251)$	& 2 & 100 & 0.42 & 6.083 & 0[\textbf{216}] & (55, 55, 0)\\
	\midrule
	$14\times 8$ & $4,111,631$ & 1 & 200 & 0.39 & 16.167 & 6.083 & (164, 178, 136)\\
	& $(16381\times 251)$	& 2 & 200 & 0.44 & 6.083 & 0[\textbf{467}] & (136, 136, 0)\\
	\midrule
	 &  & 1 & 1 & 0.4 & 20.333 & 12.167   & (279, 126, 217) \\
	& 8,219,999	& 2 & 100 & 0.43 & 12.167 & 8.000 & (217, 180, 277)\\
	$15\times 8$ & $(32749\times 251)$	& 3 & 200 & 0.43 & 8.000 & 6.000      & (277, 65, 282)\\
	&	& 4 & 200 & 0.44 & 6.000 & 4.083  & (282, 247, 71)\\
	&	& 5 & 200 & 0.44 & 4.083 & 0[\textbf{329}] & (71, 71, 0)\\
	\bottomrule
	\end{tabular}
        \captionsetup{labelformat=parens,labelsep=space}
    \subcaption{\label{tab: more_effient_IRV} Results of the IRV variant that focuses on longer pauses.}
  \end{minipage}
  \caption{Result about IRV. The label $\Delta HAM$ reports the Hamming distance between the forward annealing lowest energy sample and the reverse annealing lowest energy sample. The labels $HAM$ and $HAM_{new}$ reports the Hamming distance of respectively the starting point and the lowest energy sample of that iteration with respect to the ground state. The bold number near 0 reports how many samples reached 0 energy for that iteration, out of 1000.}  
\end{table}

\ignore{
\begin{table}[t]
  \begin{minipage}[b]{1\linewidth}  
    \centering
    \scriptsize
    \begin{tabular}{|c|c|c|c|c|rr|rr|r|}
	\toprule
	size & $P\ (A\times B)$ & \# & $T_p$ & $S_p$ & $min(P_F)$ & $min(P_F)_{new}$  & $Ham$ & $Ham_{new}$ & $\Delta Ham$ \\
	\midrule
	 & $1,027,343$  & 1 & 1 & 0.31 & 10.167 & 4.000 & 263 & 151 & 142\\
	$12\times 8$ & ($4093\times 251$)	& 2 & 1 & 0.38 & 10.167 & 4.083 & 128 & 58 & 122\\
	&	& 3 & 100 & 0.38 & 4.083 & 0[\textbf{2}] & 58 & 0 & 58\\
	\midrule
	 &  & 1 & 1 & 0.35 & 18.167 & 8.083 & 290 & 249 & 353 \\
	$14\times 8$ & $4,111,631$	& 2 & 1 & 035 & 16.000 & 6.000 			& 273 & 240 & 219 \\
	& $(16381\times 251)$	& 3 & 50 & 0.37 & 10.000 & 2.083 	& 277 & 85 & 280\\
	&	& 4 & 10 & 0.38 & 2.083 & 0[\textbf{67}]  & 85 & 0 & 85\\
	\midrule
	 & & 1 & 1 & 0.34 & 18.333 & 6.083 & (, 294,) \\
	$16\times 8$ & 16,445,771	& 2 & 10 & 0.35 & 10.000 & 4.000 		& & & 374\\
	& $(251 \times 65521)$	& 3 & 50 & 0.36 & 6.083 & 4.167 	& & & 292\\
	&	& 4 & 100 & 0.39 & 4.167 & 4.000		& & & \\
	\bottomrule
	\end{tabular}
     \captionsetup{labelformat=parens,labelsep=space}
    \subcaption{\label{tab: IRV} Results of the original IRV algorithm. The bold number near 0 reports how many samples reached 0 energy for that iteration, out of 1000.}
  \end{minipage}
  \begin{minipage}[b]{1\linewidth}  
    \centering
    \scriptsize
    \begin{tabular}{|l|c|c|c|c|rr|c|}
	\toprule
	size & $P(A\times B)$ & \# & $T_p$ & $S_p$ & $min(P_F)$ &   $min(P_F)_{new}$ & $(HAM, \Delta HAM, HAM_{new})$ \\
	\midrule
	$13\times 8$ & $2,055,941$ & 1 & 100 & 0.38 & 14.083 & 6.083 & (204, 185, 55) \\
	& $(8191\times 251)$	& 2 & 100 & 0.42 & 6.083 & 0[\textbf{216}] & (55, 55, 0)\\
	\midrule
	$14\times 8$ & $4,111,631$ & 1 & 200 & 0.39 & 16.167 & 6.083 & (164, 178, 136)\\
	& $(16381\times 251)$	& 2 & 200 & 0.44 & 6.083 & 0[\textbf{467}] & (136, 136, 0)\\
	\midrule
	 &  & 1 & 1 & 0.4 & 20.333 & 12.167   & (279, 126, 217) \\
	& 8,219,999	& 2 & 100 & 0.43 & 12.167 & 8.000 & (217, 180, 277)\\
	$15\times 8$ & $(32749\times 251)$	& 3 & 200 & 0.43 & 8.000 & 6.000      & (277, 65, 282)\\
	&	& 4 & 200 & 0.44 & 6.000 & 4.083  & (282, 247, 71)\\
	&	& 5 & 200 & 0.44 & 4.083 & 0[\textbf{329}] & (71, 71, 0)\\
	\bottomrule
	\end{tabular}
        \captionsetup{labelformat=parens,labelsep=space}
    \subcaption{\label{tab: more_effient_IRV} Results of the IRV variant that focuses on reducing the number of iterations. The bold number near 0 reports how many samples reached 0 energy for that iteration, out of 1000.}
  \end{minipage}
  \caption{Result about iterative reverse annealing.}  
\end{table}
}

Overall, exploiting all the encoding and solving techniques described
in this paper, 
 $8,219,999=32,749\!\times\!251$ was the highest prime product we were able to factorize.
  To the best of our knowledge, this is the largest number that
  was ever factorized by means of quantum annealing.